 \documentclass[preprint,12pt]{aastex}
 \usepackage{epstopdf}

\shorttitle{UCD Wide Common Proper Motion Pairs}
\shortauthors{Faherty et al.}

\begin{document}


\title{The Brown Dwarf Kinematics Project (BDKP). II. Details on Nine Wide Common Proper Motion Very Low--Mass Companions to Nearby Stars\footnote{This paper includes data gathered with the 6.5 meter Magellan Telescopes located at Las Campanas Observatory, Chile.} \footnote{This paper makes use of data obtained with the telescopes operated by the SMARTS consortium}}


\author{Jacqueline K. Faherty\altaffilmark{1,3,6}, Adam J.\ Burgasser\altaffilmark{2,7}, Andrew A.\  West\altaffilmark{2}, John J.\  Bochanski\altaffilmark{2},  Kelle L.\ Cruz\altaffilmark{4,5}, Michael M. Shara\altaffilmark{1}, Frederick M. Walter\altaffilmark{3,8}}

\altaffiltext{1}{Department of Astrophysics, 
American Museum of Natural History, Central Park West at 79th Street, New York, NY 10034; jfaherty@amnh.org }
\altaffiltext{2}{Massachusetts Institute of Technology, Kavli Institute for Astrophysics and Space Research,
Building 37,  77 Massachusetts Avenue, Cambridge, MA 02139}
\altaffiltext{3}{Department of Physics and Astronomy, Stony Brook University Stony Brook, NY 11794-3800}
\altaffiltext{4}{Astronomy Department, California Institute of Technology, Pasadena, CA 91125}
\altaffiltext{5}{Spitzer Postdoctoral Fellow}
\altaffiltext{6}{Visiting Astronomer at the Infrared Telescope Facility, which is operated by the University of Hawaii under Cooperative Agreement no. NCC 5-538 with the National Aeronautics and Space Administration, Science Mission Directorate, Planetary Astronomy Program.}
\altaffiltext{7}{Center of Astrophysics and Space Sciences, Department of Physics, University of California, San Diego, CA 92093, USA}
\altaffiltext{8}{Guest observer at the Kitt Peak National Observatory, which is operated by the Association of Universities  for Research in Astronomy under cooperative agreement with the National Science Foundation}

\begin{abstract}
We report on nine wide common proper motion systems containing late-type M, L, or T companions.  We confirm six previously reported companions, and identify three new systems.  The ages of these systems are determined using diagnostics for both stellar primaries and low--mass secondaries and masses for the secondaries are inferred using evolutionary models.  Of our three new discoveries, the M3+T6.5 pair G 204-39 and SDSS J1758+4633 has an age constrained to 0.5-1.5 Gyr making the secondary a potentially useful brown dwarf benchmark. The G5+L4 pair G 200-28 and SDSS J1416+5006 has a projected separation of $\sim$25,000 AU making it one of the widest and lowest binding energy systems known to date.  The system containing NLTT 2274 and SDSS J0041+1341 is an older M4+L0 ($>$4.5 Gyr) pair which shows H$\alpha$ activity in the secondary but not the primary making it a useful tracer of age/mass/activity trends. Two of the nine systems have discrepant component ages which emerge from stellar or ultracool diagnostics indicating possible shortcomings in our understanding of the age diagnostics of stars and brown dwarfs. We find a resolved binary frequency for widely-separated ($>$ 100 AU) low--mass companions (i.e. at least a triple system)  which is at least twice the frequency found for the field ultracool dwarf population.  The ratio of triples to binaries and quadruples to binaries is also high for this sample: 3:5 and 1:4, respectively, compared to 8-parsec sample values of 1:4 and 1:26.  The additional components in these wide companion systems indicates a formation mechanism that requires a third or fourth component to maintain gravitational stability or facilitate the exchange of angular momentum.  The binding energies for the nine multiples discussed in this text are among the lowest known for wide low-mass systems, suggesting that weakly bound, low--to--intermediate mass (0.2M$_{\sun}$ $<$ M$_{tot}$ $<$1.0M$_{\sun}$) multiples can form and survive to exist in the field (1-8 Gyr).  


\end{abstract}

\section{INTRODUCTION}
Ultracool dwarfs (UCDs) comprise the late-type M, L, and T dwarf spectral classifications  \citep[e.g.,][and references therein]{2005ARA&A..43..195K} and include brown dwarfs--objects that do not support stable hydrogen fusion  (\citealt{1962AJ.....67S.579K}; \citealt{1963PThPh..30..460H}).  UCDs sample the low-mass extremum of star formation processes and are abundant in nearly every Galactic environment.  The low temperatures and high pressures in the photospheres of UCDs give rise to abundant molecular species, whose complex chemistry and opacities result in highly-structured spectral energy distributions (SEDs).  Disentangling the physical characteristics---mass, age, surface gravity, metallicity, atmospheric properties, etc.--- that modulate these spectral features is a critical step for testing theoretical models.  However, individual characterization of Galactic brown dwarfs is challenging because their thermal evolution leads to a degeneracy between mass, age, and physical properties derived from observables such as luminosity and effective temperature (T$_{eff}$).  While spectral analyses can constrain physical properties for some systems (e.g. \citealt{2006ApJ...639.1095B}; \citealt{2007ApJ...656.1136S}; \citealt{2007MNRAS.381.1400W}; \citealt{2008ApJ...678.1372C}), calibration of these techniques requires detailed studies of well-understood benchmark systems.

One useful group of UCD benchmarks are those which are resolved companions to nearby, well-characterized stars.  Assuming coevality, the physical properties of the primary, such as metallicity and age---which are extremely difficult to measure for low-mass stars---can be applied to the companion.  In particular, independent age determinations are critical to break the mass/age/observable degeneracy for the brown dwarf companion.  Despite the apparent scarcity of wide UCD companions to nearby stars  ($\sim$2-3\%; \citealt{2001ApJ...551L.163G}, \citealt{2008ApJ...683..844L}), several have been identified and used to calibrate spectral analysis techniques (e.g., \citealt{2006ApJ...639.1095B}; \citealt{2007ApJ...656.1136S}), as well as to critically test atmospheric (e.g., \citealt{2008ApJ...682.1256L}) and structure/evolutionary models (e.g.,  \citealt{2004ApJ...609..885M};  \citealt{2008arXiv0807.2450D}). The frequency and characteristics of widely-separated stellar-UCD pairs also puts important constraints on the star formation processes and the subsequent dynamical evolution of stellar systems (e.g., \citealt{2003ApJ...586..512B}; \citealt{2003ApJ...587..407C,2007ApJ...660.1492C}; \citealt{2007ApJ...668..492A}; \citealt{2009ApJ...691.1265L}).  However, the known population of UCD companions remains small and does not yet fully sample the range of ages, masses and metallicities found among unassociated field sources.


In the past decade, multiplicity surveys focused on the field UCD population have distinguished two classes:  
\begin{itemize}
\item  
Roughly 10-20\% of the field UCDs are found to be closely-separated ($\rho$ $<$ 20 AU), near-equal mass, small total mass (M$_{tot}<$0.2) UCD-UCD multiples (e.g. \citealt{2003AJ....126.1526B}, \citealt{2003ApJ...587..407C}, \citealt{2003ApJ...586..512B}, \citealt{2007ApJ...671.2074A}, \citealt{2008AJ....135..580R})
\item
A smaller fraction are found to be widely-separated ($\rho$ $>$ 100AU) from a much more massive stellar companion (e.g. \citealt{2001AJ....121.3235K}, \citealt{2001AJ....122.1989W}, \citealt{2005ApJ...625..385A})
\end{itemize}

In the first case, the typically tight separations for UCD binaries is well-established (e.g. Allen et. al\ 2007), and early studies by  \citet{2003ApJ...586..512B}, and \citet{2003ApJ...587..407C} identified a maximum separation limit/minimum binding energy for field UCD-UCD pairs of E$_{b} \sim$ 2$\times$10$^{42}$ erg. However, the recent discovery of a number of young UCD systems (ages $<$10 Myr) and a handful of field systems that are more widely-separated ($\rho$ $>$ 100 AU) and more weakly bound (E$_{b}$ $<<$ 10$^{42}$ erg), questions whether separation limits can be considered constraints for formation models or if wide UCD binaries are a normal (albeit rare) mechanism of UCD formation (\citealt{2005ApJ...633..452K, 2006ApJ...649..306K}, \citealt{2007ApJ...663..394K}, \citealt{2009ApJ...691.1265L}, and \citealt{2009arXiv0902.4742A}, \citealt{2007ApJ...659L..49A}, \citealt{2005A&A...440L..55B}, \citealt{2005A&A...439L..19P}, \citealt{2007ApJ...667..520C}, \citealt{2009arXiv0903.3251R}).  

In contrast, systems in the second category ($\rho$ $>>$ 100 AU) have binding energies that are several orders of magnitude smaller than the minimum set for UCD-UCD pairs. \citet{2005AJ....129.2849B} noted a higher binary frequency among UCDs that are widely-separated from a stellar primary, suggesting the need for higher masses or an angular momentum sink in multibody interactions to form these systems. Recent work by \citet{2006A&A...458..817W} suggests that for a low-mass primary fragment formed in the cooler outer parts of a circumstellar disk ($\rho$ $>$ 100 AU), and spinning at a fast enough rate, H$_{2}$ dissociation is likely to trigger a Secondary Fragmentation phase, thereby potentially giving rise to a closely-separated (a$\sim$5 AU (m$_{system}$/0.1M$_{\sun}$) UCD binary.

Current observational evidence suggests that widely-separated stellar companions exist out to distances of $\sim$0.1pc (\citealt{1984ApJ...281L..41L}; \citealt{1987ApJ...312..367W}).  Beyond this separation, perturbations from passing stars and giant molecular clouds will likely disrupt the companions over the lifetime of the Galaxy.   Separations of stellar-UCD and, especially, UCD-UCD multiples appear to fall well below the perturbation limit, suggesting dynamical sculpting occurs only in the natal environment (\citealt{2003ApJ...586..512B}; \citealt{2007ApJ...660.1492C}). However, the current sample of such systems is far from complete.  In large part this is due to the challenge of covering a large area of the sky, and ascertaining evidence for companionship between two objects.  For stars, common proper motions have been the standard characteristic for identifying co-moving objects at large angular separations (\cite{1961AJ.....66..528V,1944AJ.....51...61V}; \citealt{1979lccs.book.....L}; \citealt{2002AJ....124.1190L}).  Historically, optical proper motion catalogs lacked the depth to detect late-type M, L, and T dwarfs. In addition, the recent discovery of UCDs has largely precluded astrometric measurements due to short temporal baselines, making an extensive common proper motion search difficult. In the past few years, large UCD proper motion samples (e.g., \citealt{2008MNRAS.384.1399J}; \citealt{2008MNRAS.390.1517C}; \citealt{2009AJ....137....1F}) and near-IR proper motion surveys have become available (e.g., \citealt{2005A&A...435..363D,2007A&A...468..163D,2009MNRAS.394..857D}), making it possible to perform a search in the reverse direction: using the UCD proper motion to find a stellar companion.

In this study, we used a proper motion catalog of UCDs from \citet{2009AJ....137....1F} (hereafter, the BDKP catalog) to conduct a common proper motion search for main sequence companions to Hipparcos  (\citealt{1997A&A...323L..49P}) or LSPM-N (\citealt{2002AJ....124.1190L}) catalog stars.  We have uncovered nine systems, six of which have been briefly noted in the literature and three of which are presented here for the first time.   In section 2 we discuss our target list, the criteria for companionship and the reliability of our matches.  In section 3 we discuss follow-up photometry as well as optical and near-IR spectroscopy of our candidate systems.  In section 4 we apply age diagnostic tests to the primaries and secondaries and calculate masses of the UCD secondaries. In section 5 we explore the stability of the nine systems as well as multiplicity and formation mechanisms for a large sample of UCD field companions.  We summarize our results in section 6.

\section{WIDE COMPANION DISCOVERY}
\subsection{Initial Target List and Selection Criteria}
We began an astrometric search for common proper motion candidates to UCDs using the BDKP catalog (\citealt{2009AJ....137....1F}) of 842 late-type M, L, and T dwarfs.  The catalog is composed of 570 L and T dwarfs (all of which can be found on the DwarfArchives compendium\footnote{http://dwarfarchives.org}) and 272 M7-M9 dwarfs drawn from the literature.  Objects span spectral types from M7-T8 and cover a wide range of magnitudes, distances, and proper motions. 

To avoid a large number of chance alignments with slowly moving objects,  we only considered the 681 UCDs in the BDKP catalog with proper motion $>$ 100 mas yr$^{-1}$.  We compared the positions and motions of the UCDs to stars in the Hipparcos (\citealt{1997A&A...323L..49P}) and LSPM-N (\citealt{2002AJ....124.1190L}) catalogs.  An angular separation of up to 10 arcminutes and a proper motion match criterion of better than 2$\sigma$ in both RA and DEC  were required between the system components. The average uncertainty for objects in the BDKP catalog is 15 mas yr$^{-1}$ so we typically required an agreement in proper motion $<$ 30 mas yr$^{-1}$ between the stellar companion and UCD. 

We also used distances to further rule out chance alignment pairs.  All of the UCDs listed in the BDKP catalog have photometric distance estimates based on the \citet{2003AJ....126.2421C} relation for M7-L5 dwarfs or the \citet{2007ApJ...659..655B} relation for L6-T8 dwarfs.  All of the stellar candidate companions had either photometric distances of their own (\citealt{2005AJ....130.1680L}) or had parallax measurements from the Hipparcos catalog.  We required a distance agreement of better than 2$\sigma$, which generally meant $<$ 10 pc difference.   


\subsection{New Candidate Companion Systems}

After selecting by angular separation, proper motion, and distance we were left with 30 possible wide common proper motion pairs with a Hipparcos or LSPM-N star.   Twenty-one of these were previously known systems and are listed in Table~\ref{All} and not discussed at length within this study.  Six systems with UCD components: 2MASS J0003-2822, 2MASS J0025+4759, SDSS J0041+1341, SDSS J0207+1355, 2MASS J1320+0957, and 2MASS J1320+0409, have been previously reported in the literature but not studied in detail ( \citealt{2003AJ....126.2421C}, \citealt{2006MNRAS.368.1281P},  \citealt{2008MNRAS.384.1399J}, \citealt{2009MNRAS.394..857D}).  Three systems with UCD components  2MASS J1200+2048, 2MASS J1416+5006, and SDSS J1758+4633, are reported here for the first time. These nine systems are summarized in Table~\ref{Astro_Comp}.  


\subsection{Reliability of Common Proper Motion Candidates}

\begin{figure*}[!ht]
\begin{center}
\plottwo{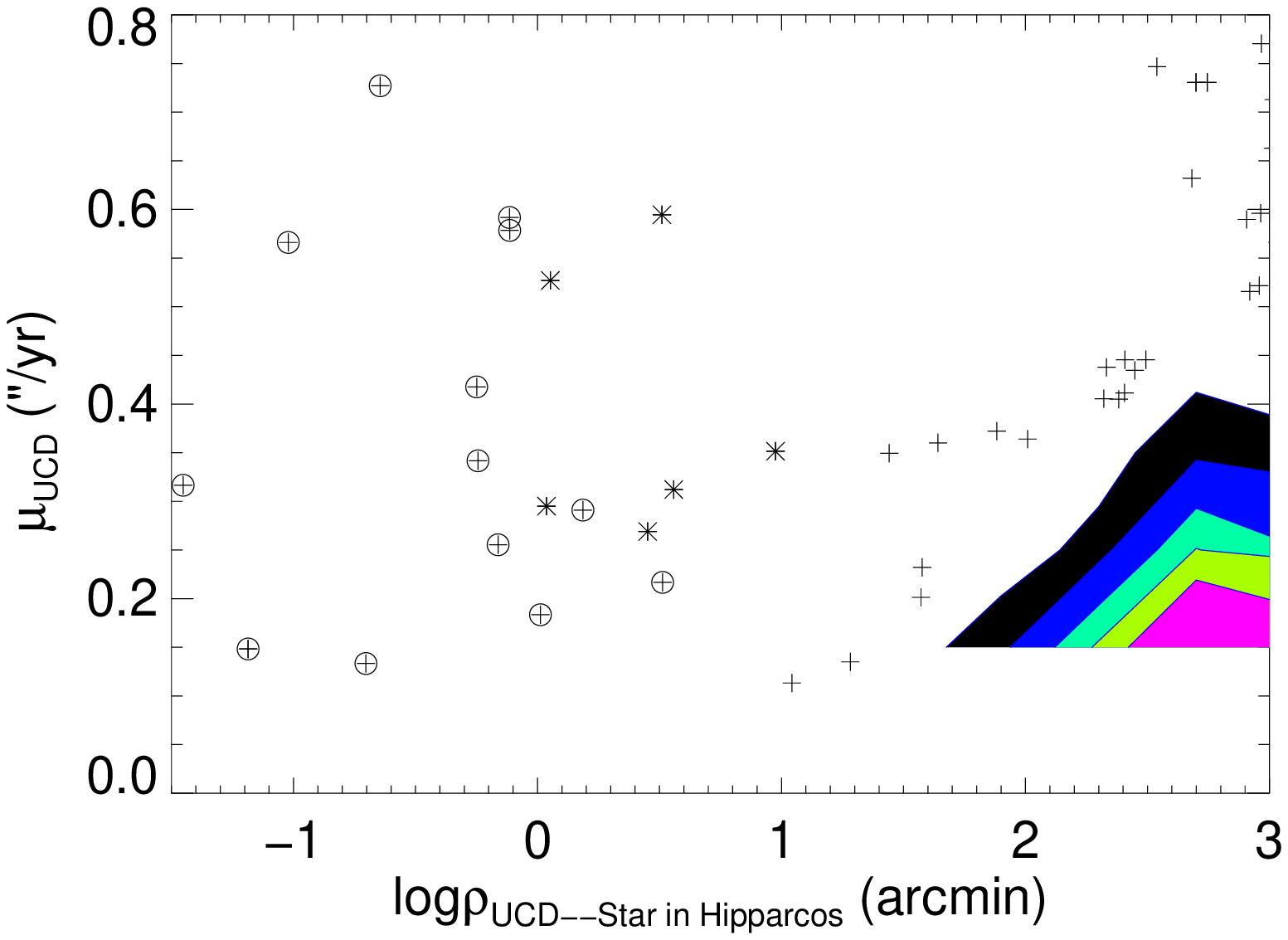}{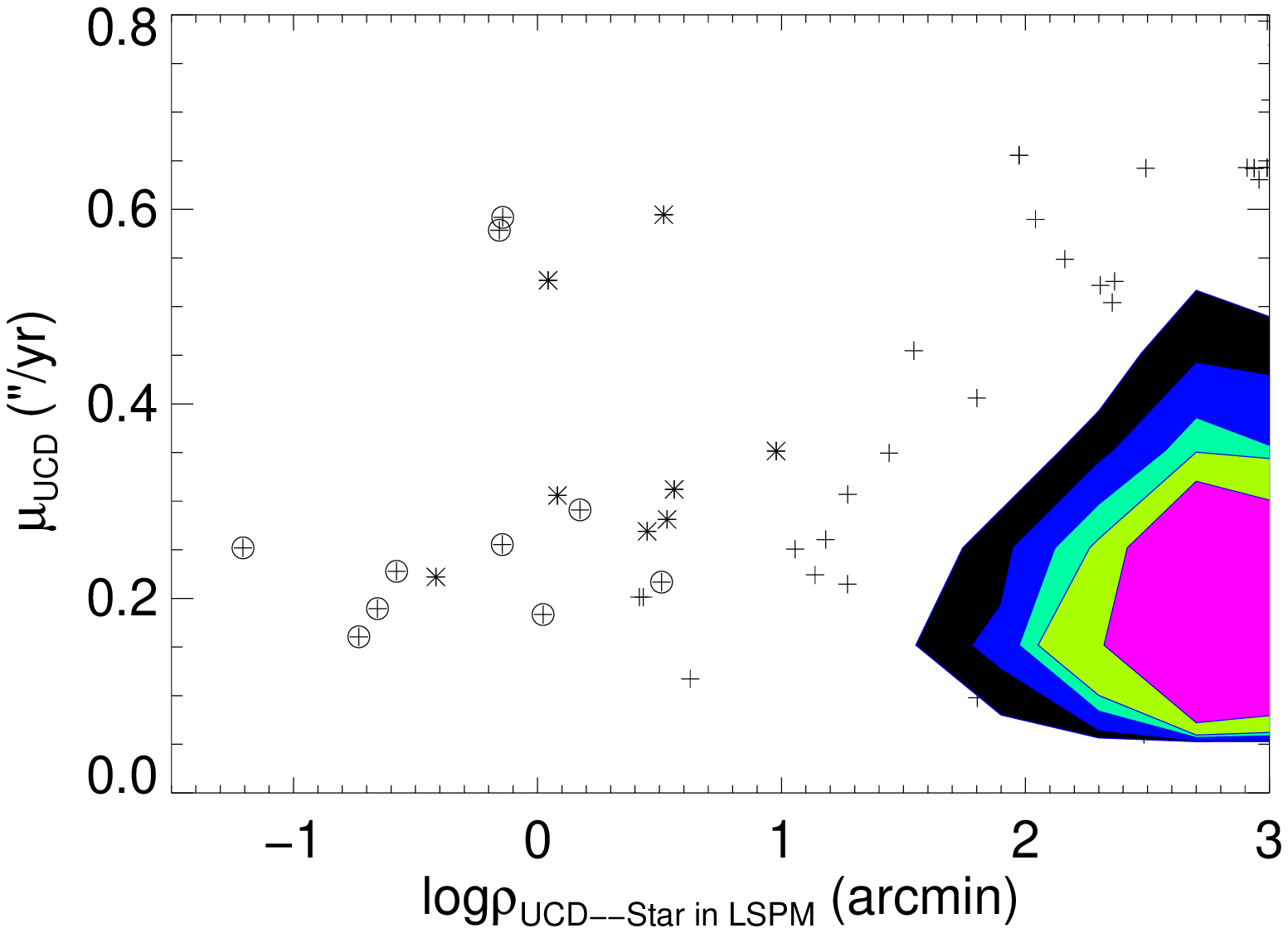}
\end{center}
\caption{ Proper motion vs. separation of the known and potential common proper motion pairs of Hipparcos stars  (left panel) and LSPM-North stars (right panel)  to UCDs in the BDKP catalog moving faster than 100 mas yr$^{-1}$.  We required a proper motion component match of 2$\sigma$ between star and UCD.  There was no distance requirement between potential pairs applied in this plot.  Objects marked by circles are previously published wide ultracool dwarf pairs.  Objects marked by asterisks are wide ultracool pairs discussed in this paper.  In the right plot, we rejected two objects within the 10 arc-minute radius because their photometric distances were greater than 3$\sigma$ from the UCD.  The contours in each plot represent densities of  75, 200, 500, 750, and 2000 objects.} 
\label{fig:HIP_LSPM_Match}
\end{figure*}

To quantify the probability that our pairs might be chance alignments, we ran a Monte Carlo simulation of all stars in the LSPM-N and Hipparcos catalogs that shared a common proper motion, but not necessarily distance or position, with our UCDs (to within 2$\sigma$).  We assumed that high proper motion objects are rare so we can accurately sample the observed proper motion distributions in the LSPM-N and Hipparcos catalogs.  For computational purposes we created a simulation grid that was equal in angular size to the area covered by the catalogs. LSPM-N is over 99$\%$ complete at high galactic latitudes and over 90$\%$ complete at low galactic latitudes so we assume an area of half the sky for this survey. Hipparcos is an all-sky catalog and depending on galactic latitude and spectral type, complete to V$\sim$7.3-9.0. The resolution of each grid point was set to be the angular separation between the pairs discussed in Table~\ref{Astro_Comp}.   Our simulation drew N stars (where N is the number of stars with matching proper motions) and placed them randomly in the grid.   The number of times two stars fell in the same grid region (or within the observed pair separation) was determined. We iterated each simulation 10000, 100000 or 1000000 times, depending on the iterations required to produce a chance alignment.   The ratio of matches to trials provided a probability for random association, as listed in Table~\ref{Reliability}.  The simulations are based solely on the distributions of proper motions in empirical data and do not account for the spatial distribution of the stars on the sky or any models of Galactic structure, both of which would likely decrease the probability of chance alignment.  We found that the likelihood that any of the nine systems in Table~\ref{Astro_Comp} is a chance coincidence is  $<$ 0.01\%.  Figure~\ref{fig:HIP_LSPM_Match} illustrates the reliability of the new common proper motion pairs.  We investigated the spatial distribution of these matches and found no preferred direction indicating that the matches are indeed randomly selected.  Only two objects within a 10 arcminute separation did not have matching distances (see the LSPM-N matches in the right panel of Figure~\ref{fig:HIP_LSPM_Match}).

\citet{2007AJ....133..889L} performed a similar proper motion reliability check by comparing the entire LSPM-N catalog to the Hipparcos catalog.  They used over 4000 known Hipparcos stars that had a wide LSPM-N companion and then simulated chance alignments in those fields by moving from 1 to 5 degrees away from the known pair and evaluating any additional systems that shared the same proper motion.  They derived the following relation which is globally applicable for any pair of co-moving stars with $\mu>$0.15$\arcsec$/yr and tests whether a common proper motion system has $>$50\% probability of being physically associated: 
\begin{equation}
\Delta\theta\Delta\mu<(\mu/0.15)^{3.8}
\end{equation}

\noindent where $\Delta\theta$ is the angular separation (in $\arcsec$), $\mu$ is the mean total proper motion of the pair in $\arcsec$ yr$^{-1}$, and $\Delta\mu$ is the magnitude of the difference between the proper motion vectors in $\arcsec$ yr$^{-1}$. 

We have used this relation as a second reliability check on each of our pairs and find that the nine systems from Table~\ref{Astro_Comp} pass this criterion.  

\section{OBSERVATIONS}
\subsection{Optical Spectroscopy with SMARTS}
\subsubsection{R-C Spectrograph}
Optical spectra for six of the primaries were obtained with the R-C spectrograph on the CTIO SMARTS 1.5m telescope over several nights in the fall of 2008 and winter of 2009. Table~\ref{SMARTS} provides details of our observations.  The R-C is a slit spectrograph, with a 300$\arcsec$ long slit oriented east-west.   We employed various spectral setups that covered either the red or blue part of the spectrum (see Table~\ref{SMARTS} for details).  The detector is a Loral 1K CCD with 1199 pixels in the direction of the dispersion.  All spectra were acquired through queue observing with time allocated through the SMARTS consortium.  The conditions for these observations were moderate with an average seeing of 1.0 - 1.2 $\arcsec$. Targets were observed through a 110$\mu$m (2.0$\arcsec$) wide slit. Three images of each target were obtained and accompanied by a wavelength calibration exposure of a Ne-Ar or Th-Ar arc lamp. A spectro-photometric standard, either Feige 110 or LTT 4364, was observed each night for flux calibration.  Images were bias-subtracted, trimmed, and flattened, then co-added using a median filter. Spectra were extracted using IDL routines that fit a Gaussian in the spatial dimension at each column in the CCD. The net counts at each pixel are the integrated counts in the Gaussian, less the interpolated background fit to either side of the spectrum.  

\subsubsection{Echelle Spectrograph}
High dispersion spectra of three of the primaries (Table~\ref{SMARTS} ) were obtained with the bench-echelle spectrograph on the CTIO SMARTS 1.5m telescope over three nights in the fall of 2008 and winter of 2009. Formerly mounted at the Cassegrain focus of the Blanco 4m telescope, the echelle spectrograph is fiber-fed from the 1.5m and uses a 31.6 line/mm echelle and a 226 line/mm cross disperser feeding a 2K SITE CCD detector. Our observations employed a 60~$\mu$m slit which corresponds to a 2 pixel resolution of R$\sim$40,000.  All spectra were acquired through queue observing.  The conditions for these observations were moderate with an average seeing of 1.0 - 1.2 $\arcsec$.  A quartz lamp exposure at the start of the night was obtained for flat fielding. Three images of 1500 s were obtained for each target followed by a wavelength calibration exposure of Th-Ar.  The data were reduced using IDL routines.   We  median filtered and co-added the flat field and science spectra for each target.  Using the quartz lamp trace we extracted individual spectra, and then divided by the extracted flat field spectra. The Th-Ar spectra were cross-correlated against a template spectrum to  determine the pixel shifts. The wavelength stability of the system is better than 0.5~km s$^{-1}$ over the course of half a year. The extracted spectra were linearized using the wavelength solution. Our detection equivalent width for atomic absorption features, in a 1 hour exposure at V$\sim$9, is 3~m\AA.

\subsection{KPNO Echelle Spectroscopy}
High dispersion spectra of four of the primaries (Table ~\ref{KPNO}) were obtained with the KPNO 4.0m echelle spectrograph during the nights of 2008 June 25-29 (UT). We used the 58.5 echelle grating, the 226-1 cross-disperser in second order, and the CuSO$_4$ filter to obtain spectra between about 3700 and 5000\AA.  The weather conditions for these observations were poor with an average seeing of 1-2~$\arcsec$. The rapidly changing sky conditions precluded precise focussing, and required hand-guiding.  We observed with a 1~$\arcsec$ slit and a 9.73~$\arcsec$ decker.  A ThAr lamp spectrum was obtained at each telescope position for wavelength calibration. At the start of the night, we observed the pflat lamp though a 15~$\arcsec$ decker.   Data extraction used conventional techniques. The bias was subtracted from the science frame which was then divided by the lengthened flat. Targets were self-traced during extraction and the background was estimated from the region above and below the target on the slit.  For the primary G~62-33, a weighted sum of the two spectra taken on 2008 June 26 and 27 was used to improve S/N.  The reciprocal dispersion in the order containing Ca~II K\&H is 0.05A/pixel and the nominal instrumental resolution is R$\sim$33,000.  

We followed the technique used by \citet{1979ApJS...41..481L} to directly measure R$^{\prime}_{HK}$ from the echelle data.  First we normalized the spectrum by scaling it to a flux-calibrated low-dispersion spectrum. This removed any residual instrumental signature remaining after flattening the spectrum. Then we scaled to an absolute surface flux using Linsky's calibration of \citet{1965MmRAS..69...83W} photometry in the 3925-3975\AA\ bandpass. This calibration uses Johnson $V-R$ colors so we converted the $B-V$ colors to Cousins $V-R_C$, and then used the transformation in \citet{1979PASP...91..589B} to convert to $V-R$. We measured the flux between the K$_1$ and H$_1$ minima and interpolated the photospheric contribution to the flux between them using the data in \citet{1979ApJS...41..481L}. R$^{\prime}_{HK}$ is the net surface flux normalized to $\sigma$T$_{eff}^4$. 

We verified the technique by measuring R$^{\prime}_{HK}$ for 5 calibration stars: $\xi$~Boo~A,B, 61~Cyg~A,B, and HD 128165. With the exception of $\xi$~Boo~B which was high by$\sim$50\%, all measurements agreed with published values to within 10-20\%.  We note that H$\epsilon$ is seen prominently in emission in the spectrum of $\xi$~Boo~B, so the star was likely flaring. Examination of chromospheric emission levels in \citet{1995ApJ...438..269B} shows that variations of 10-50\% are common over the course of stellar magnetic cycles. We also measured the solar (twilight sky) spectrum and calculated the solar log(R$^{\prime}_{HK}$) = -4.8$^{+0.2}_{-0.3}$.

\subsection{Optical Spectroscopy with MagE}
Optical spectra for five of the primaries and three of the UCD secondaries were obtained with the Magellan Echellette Spectrograph (MAGE; \citealt{2008SPIE.7014E.169M}) on the 6.5m Clay Telescope at Las Campanas Observatory over several nights in October 2008, November 2008, and January 2009.  Table~\ref{MagE} lists the details of our observations.  MagE is a cross--dispersed optical spectrograph, covering 3,000 to 10,000~\AA~ at medium resolution ($R \sim 4,100$).  Our observations employed a $0.7^{\prime\prime}$ slit aligned at the parallactic angle, and the chip was unbinned.   These observations were made under clear conditions with an average seeing of $\sim$0.7$\arcsec$.  The targets were first acquired with the MagE finder camera using an $\it{R}$ filter.  For the primaries we used 5-30s exposures for the brightest targets and 100-120s exposures for the faintest.  For the UCD secondaries we used 1200-2400s.  A ThAr lamp spectrum was obtained at each telescope position for wavelength calibration and the spectrophotometric standard GD 108 was observed during each run for flux calibration purposes.  Ten Xe-flash and Quartz lamp flats as well as twilight flats were taken at the start of each evening for pixel response calibration.  The data were reduced using a preliminary version of the MagE Spectral Extractor pipeline (MASE; Bochanski et al.,  in prep) which incorporates flat fielding, sky subtraction and flux calibration IDL routines. 

\subsection{Near-Infrared Spectroscopy with SPEX}
Near-IR spectra for two of the primaries and three of the UCD secondaries were obtained with the SpeX spectrograph mounted on the 3m NASA Infrared Telescope Facility (IRTF) over several nights in December 2008.   The conditions of this run were variable with patchy clouds and average seeing (0.8 -1.0 $\arcsec$ at $\it{J}$).    Table~\ref{SpeX} lists the details of our observations.  We operated in prism mode with the 0.8$\arcsec$ slit aligned at the parallactic angle and obtained low-resolution ($\lambda$/$\Delta$ $\lambda$ $\sim$90) near-infrared spectral data spanning 0.7 - 2.5 $\mu$m .  Each target was first acquired in the guider camera. Exposure times varied from 120s to 150s depending on the brightness of the target.  Six images were obtained for each object in an ABBA dither pattern along the slit.   An A0V star was observed immediately after each target at a similar airmass for flux calibration and telluric correction.  Internal flat-field and Ar arc lamp exposures were acquired for pixel response and wavelength calibration, respectively.  All data were reduced using SpeXtool version 3.3 (\citealt{2003PASP..115..389V}, \citealt{2004PASP..116..362C}) using standard settings.  

\subsection{Photometric Follow-Up}
Optical photometry for four of the primaries (Table~\ref{PHOTO}) was obtained with the ANDICAM dual channel photometer on the CTIO SMARTS 1.3m telescope over several months in the winter of 2008 and spring of 2009.   The ANDICAM optical detector is a Fairchild 447 2048$\times$2048 CCD and was used in 2$\times$2 binning mode, yielding a nominal plate scale of 0.369 arcsec pixel$^{-1}$.  The $\sim$6.2 arcmin field of view allowed between 3-7 reference stars for photometric comparison in each image.  All data were taken by service (or queue) observing in $I$, $V$, and/or $B$ bands and nightly conditions varied.  Domeflats were taken at the start of each night and science frames were flat-fielded and trimmed using standard IRAF tasks prior to delivery.  Differential photometry was performed using IDL routines which utilized a 9 pixel aperture and a background annulus evaluated between 19 and 36 pixels from the target.

\section{CHARACTERIZING THE SYSTEMS}
Nearby solar-type stars are generally well-characterized with spectral type, metallicity, activity, radial velocity, distance, rotation, and other measureable diagnostic parameters.  As such, these companions serve to constrain the properties of the UCD counterparts.  The primaries discussed in this paper range in spectral type from F8-M4 and are all within 50 pc of the Sun. We combined the data available for them in the literature with follow-up spectroscopy and photometry with the primary goal of obtaining an age.   For the bright primaries, we used standard and template spectra provided within the IDL package the Hammer (\citealt{2007AJ....134.2398C})\footnote{ http://www.cfa.harvard.edu/$\sim$kcovey/thehammer.html} as well as spectral standards from the Stony Brook/SMARTS Spectral Standards Library\footnote{http://www.astro.sunysb.edu/fwalter/SMARTS/spstds.html} to characterize the stars.  For the fainter secondaries we used the M dwarf templates from  \citet{2007AJ....133..531B}; the L dwarf standards from \citet{1999ApJ...519..802K} and \citet{2000AJ....120..447K}, or data available from the SpeX Prism Libraries\footnote{http://www.browndwarfs.org/spexprism/} to characterize each source. 


\subsection{Age-Dating The Systems}
There are a number of age-dating techniques for solar analogs that can constrain ages to within a few Gyrs (\citealt{2008ApJ...687.1264M}, \citealt{1999A&A...348..897L}).  The techniques employed in this study were as follows:
\begin{itemize}
\item Gyrochronology: The ages of field stars are determined based on their rotational rates. \citet{2007ApJ...669.1167B} derived a color-dependent version of the \citet{1972ApJ...171..565S} law basing the timescale for stellar rotational decay on the Sun.  For the systems for which we have rotation periods, we derive gyro ages using the Mamajek \& Hillenbrand (2008) reformulation of Barnes' (2007) formula. The Mamajek \& Hillenbrand gyro ages are typically about a factor of two larger than those derived using Barnes' coefficients. 
\item X-ray emission: Coronal activity as traced by X-ray emission is an age diagnostic, as magnetic activity declines as a star spins down over time (e.g. \citealt{1995ApJ...450..401F}).
\item  Ca II H \& K  lines: The $R'_{HK}$ index measures the amount of chromospheric emission that arises in the cores of the Ca II H \& K lines and has been observed to decay with age (\citealt{1963ApJ...138..832W}; \citealt{1972ApJ...171..565S}; \citealt{1983ApJS...53....1S}; \citealt{1991ApJ...375..722S}). \citet{2008ApJ...687.1264M} recently revised the $R'_{HK}$ activity relation for F7-K2 dwarfs (0.5$<$B-V$<$0.9 mag) and defined the following age:
\begin{equation}
\log \tau_{1} = -38.053 - 17.912 \log R'_{HK} - 1.6675 \log(R'_{HK})^{2}
\end{equation}
\noindent where $\tau_{1}$ is in years\footnote{\citet{2008ApJ...687.1264M} also define a $\tau_{2}$ age inferred from converting the chromospheric activity levels to a rotation period via the Rossby number and then converting the rotation period to an age using the revised gyrochronology relation. We convert $\tau_{1}$ into $\tau_{2}$ ages in this text using Table 13 from the \citet{2008ApJ...687.1264M} study as these are thought to be the more representative ages.}.
\item  Lithium abundances:  Li is depleted in stellar cores early in the life of solar-type stars, so it is commonly used as an age indicator.  A comparison of Li abundances to stars in clusters with well-determined ages is likely the most appropriate usage of Li as an age diagnostic. However, as in the case with nearly all other age diagnostics, there is a large scatter in the observed EW(Li) even in coeval clusters. For field-aged stars there are few to no clusters with well-determined ages to compare to.  Therefore, for older stars, one can use the \citet{1996A&A...311..961P} NLTE curve of growth, to obtain a logarithmic depletion of Li from cosmic abundances  (log~N(Li)=3.3) and use the models of \citet{1990ApJS...74..501P} to convert this depletion into an age.  
\item  Theoretical isochrones:  Ages can be determined directly by placing stars on a theoretical HR diagram, using the observed T$_{eff}$, M$_{\it{V}}$, and [Fe/H] (\citealt{2004A&A...418..989N}).  \item   Kinematics:  While individual space motions can not be used to date objects, general information can be obtained from $\it{U,V,W}$ velocity distribution.  Studies such as \citet{1989PASP..101...54E} and \citet{1992ApJS...82..351L} have defined velocity ranges that would indicate membership in the young or old part of the galaxy.  \citet{1989AJ.....97..431E} define a U-V criterion (called the ''Eggen box") for the young disk as roughly -50 km s$^{-1}$ $<$ U $<$ +20 km s$^{-1}$ and -30 km s$^{-1}$ $<$ V$<$ 0 km s$^{-1}$ (where the convention of U positive toward the Galactic center is used).  While the age associated with membership in the young or old part of the Galaxy is uncertain, \cite{1989PASP..101...54E,1989AJ.....97..431E} roughly define the transition between the two populations as 2-3~Gyr based on the kinematic analysis of well defined cluster members (Hyades, Pleiades, NGC 752 etc.).  Admittedly, individual kinematics are a very poor age diagnostic tool and any use of space motion to age date a star needs to be viewed with caution and complimented with much more robust diagnostics.  Therefore, throughout the text we use kinematics primarily as a secondary check on other more reliable age diagnostics. 
\item Metallicity: While metallicity is an important physical property of any stellar system, it is not a reliable age indicator. \citet{2004A&A...418..989N} construct an age-metallicity diagram  for field stars, but as indicated in Figure 27 of that paper the scatter is quite large.  We cite metallicity values throughout this section as being suggestive of an older or younger age; but as with kinematics we refrain from placing a significant weight on it in our analysis. 
\end{itemize}

There are also a number of age-dating techniques for UCDs: 
\begin{itemize}
\item   Lithium absorption: In fully convective low-mass stars and higher mass brown dwarfs, primordial Li rapidly decays with age due to core fusion. \citet{1985ApJ...296..502D}, \citet{1989ApJ...345..939B}, and \citet{1998ApJ...497..253U} have shown that for masses under ~0.06 M$_{\sun}$ and ages $\gtrsim$ 500~Myr the maximum central temperature is below what is required for Lithium-burning.  This mass can be converted to an age for a given spectral type using theoretical models such as \citet{1997ASPC..119....9B}.\item   H$\alpha$ activity:  \citet{2008AJ....135..785W} suggest activity lifetimes for M0-M7 dwarfs based on H$\alpha$ equivalent width and vertical distance from the Galactic Disk Plane. \item   Surface gravity features: \citet{2007ApJ...657..511A}, \citet{2008ApJ...689.1295K}, and \citet{2009AJ....137.3345C} have shown that the presence of weak alkali spectral features, and enhanced metal oxide absorption in UCDs are best explained by lower surface gravities, implying typical ages $<$ 100 Myr. \item   $J-K_{s}$ color:   \citet{2008ApJ...689.1295K}, \citet{2008MNRAS.385.1771J}, and \citet{2009AJ....137....1F} have all shown that $J-K_{s}$ color can be used as a rough indicator of age within the UCD population.  \citet{2009AJ....137....1F} combined this with v$_{tan}$ and found that  high v$_{tan}$ objects (v$_{tan}$ $>$ 100 km s$^{-1}$)  tended  to be unusually blue for their spectral type and were considered to be older than the field population while low v$_{tan}$ objects (v$_{tan}$ $<$ 10 km s$^{-1}$) tended to be unusually red for their spectral type and were concluded to be younger than the field population (note that this metric is only indicative of an older or younger age and does not provide a direct mapping to age (however, see \citealt{2008MNRAS.385.1771J}).
\end{itemize}

   
   Age dating is fraught with large uncertainties, and some methods listed above are more reliable than others.  The analysis that follows gives details on individual systems. In Tables~\ref{Primaries} and~\ref{Secondaries} we tabulate the observational properties of the primaries and secondaries separately  to permit comparison of the age diagnostics.  In Table~\ref{Ages} we provide our adopted ages for the systems.  While we have already discussed the reliability of the common proper motion companionship in section 2, confirming similarities in the ages of the components of each system establishes the more important criterion of co-evality.

\subsection{Hipparcos Pairs}
\subsubsection{G 266-33 with 2MASS J00034227-2822410}
G 266-33 lies just over 1.1 arcminutes from 2MASS J0003-2822 and the possibility of companionship between them was first noted in \citet{2007AJ....133..439C}.  Based on our MagE spectrum, this primary is a G8 dwarf.  \citet{1996AJ....111..439H} report Ca II H \& K emission with a log $R'_{HK}$ value of -4.55.  Using the \citet{2008ApJ...687.1264M}  relation for chromospheric activity places the age of this star in the range $\tau_{2}$=0.9-1.4 Gyr.  The $\it{U,V}$ velocities fall into the Eggen Box supporting an age of $<$2 Gyr.  There are two metallicity measurements for G 266-33:  \citet{2008yCat.5128....0H} report [Fe/H]=0.07 and \citet{1998A&A...339..791R} report [Fe/H]=0.097.   The slightly metal-rich value for G 266-33 suggests a younger field age in agreement with the chromospheric and kinematic diagnostics.   The absence of Lithium absorption in the optical spectrum (W$_{\lambda}$ (Li) $<$4~m\AA,~ logN(Li)$<$-2.5) is consistent with an age older than 600 Myr.   Based on this compilation of diagnostics the age range for G 266-33 is consistent with 0.9-1.4 Gyr.

\begin{figure*}[!ht]
\begin{center}
\epsscale{1.5}
\plottwo{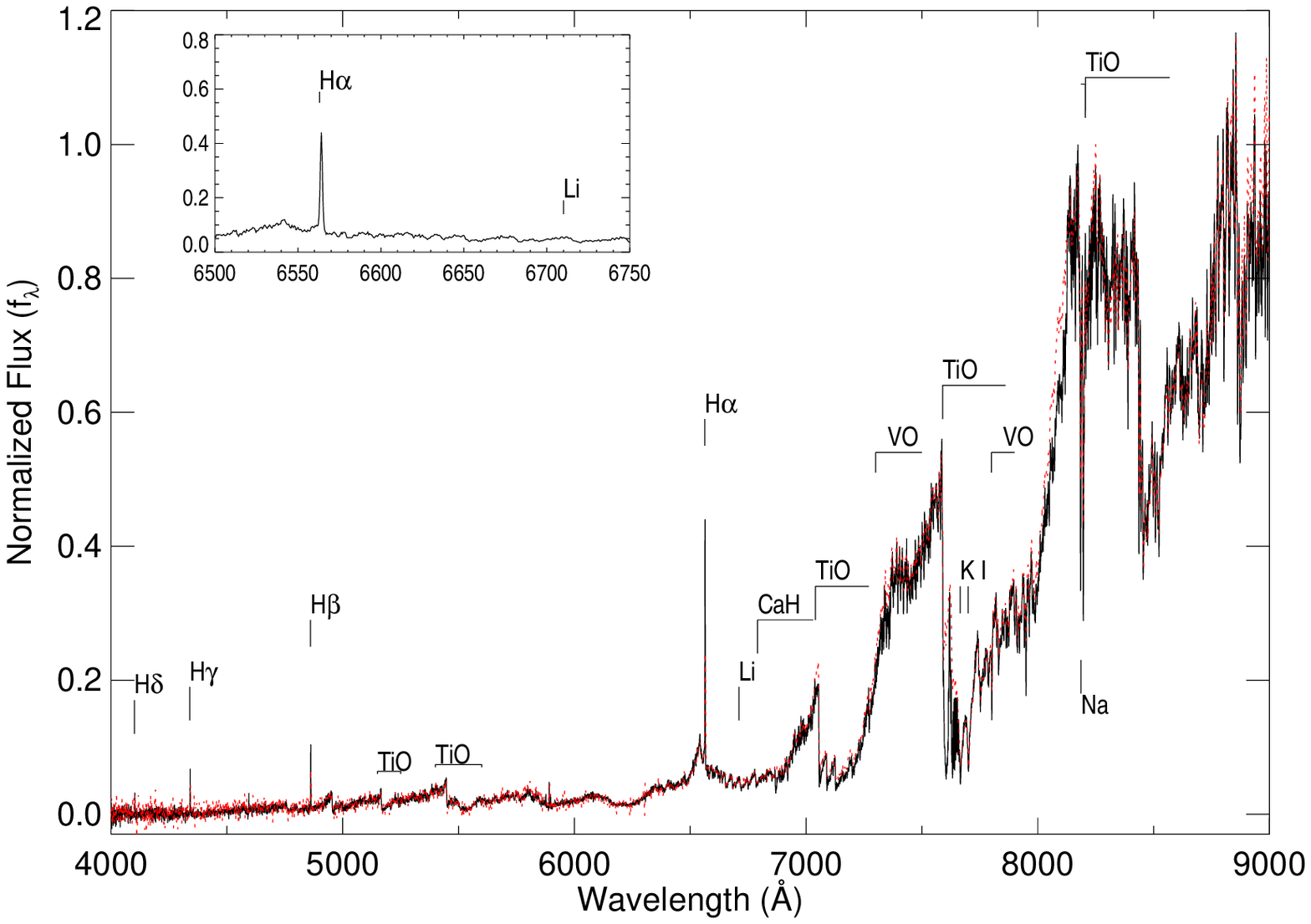}{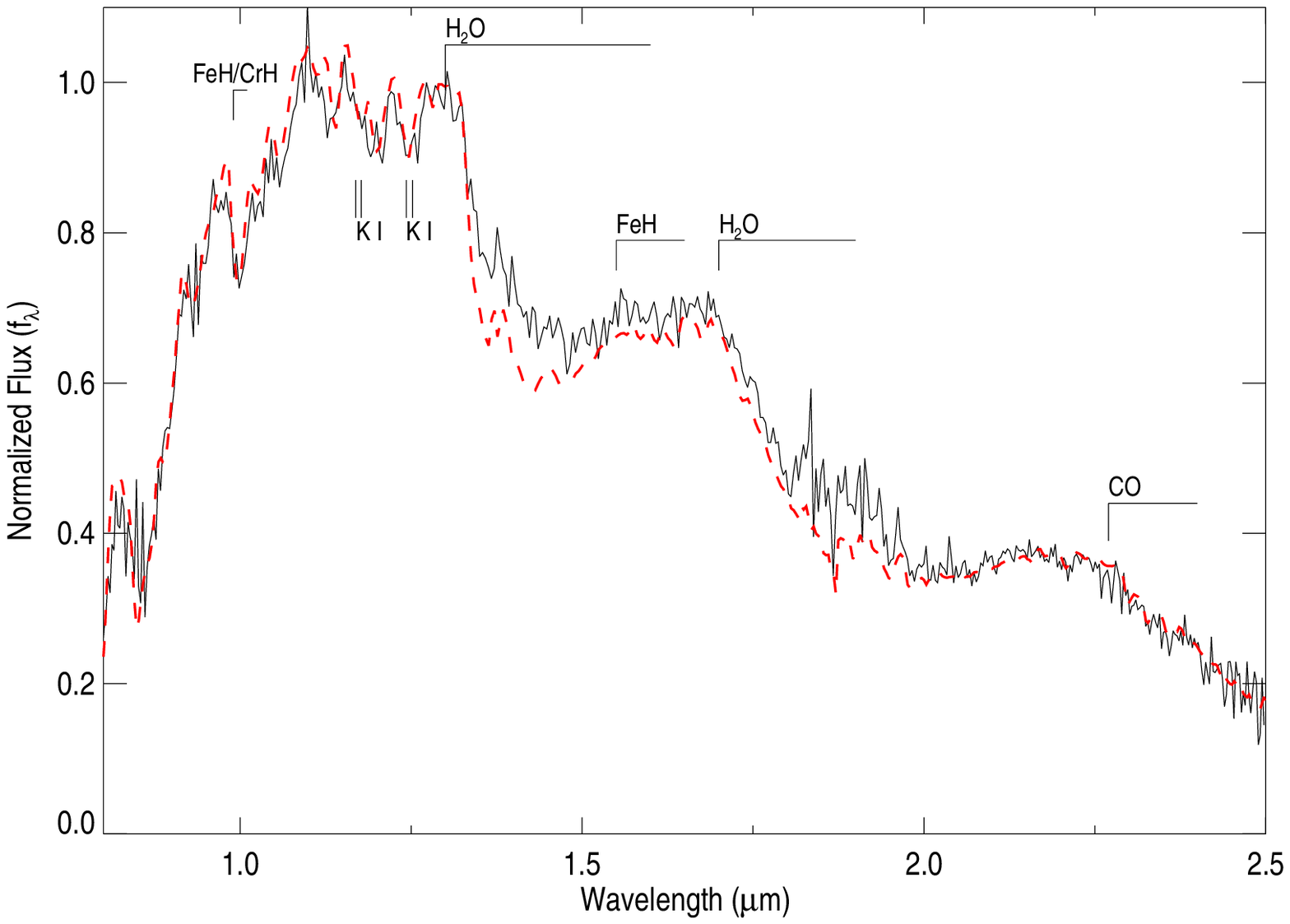}
\end{center}
\caption{The optical spectrum of the secondary 2MASSJ0003-2822 using MagE (top plot) and IR spectrum using SpeX (bottom plot).  Top:  Over-plotted is the template for an active M8 from \citet{2007AJ....133..531B} (dotted line) normalized at 8350~\AA.  The inset shows strong H$\alpha$ (6563~\AA) emission and a lack of Li (6708~\AA) absorption. Bottom: Over-plotted is the M8 optical standard VB 10 from the SpeX prism library (dotted line). } 
\label{fig:2M0003-2822}
\end{figure*}

The secondary,  2MASS J0003-2822, is classified as an M8 dwarf based on a MagE spectrum and has very strong H$\alpha$ emission, shown in Figure~\ref{fig:2M0003-2822}.  The measured H$\alpha$ equivalent width is  9.0$\pm$0.08~\AA~ and the H$\beta$, H$\delta$, and H$\gamma$ lines are also seen in emission.    For comparison, \citet{2008AJ....135..785W} examined 735 M8 dwarfs with H$\alpha$ measurements, and only 25\% of objects in that sample have stronger H$\alpha$ emission than 2MASS J0003-2822.  Combining the equivalent width of H$\alpha$ with the $\chi$ parameter from \citet{2004PASP..116.1105W} gives a log(L$_{H\alpha}$/L$_{bol}$) of -4.26.  Comparing this with other active late-type M dwarfs in \citet{West2009}, we find that 2MASS J0003-2822 is similar to the most active M7 objects (there were no M8 dwarfs for comparison).  The age determined from the age-activity relation in the \citet{West2009} study would place this object (if it were an M7) as younger than 1 Gyr.  The MagE spectrum for 2MASS J0003-2822 does not display any low-gravity features (e.g., weak Na, strong VO) and is thus likely older than 0.1 Gyr (\citealt{2008ApJ...689.1295K}).  

The $J-K_{s}$ color for 2MASS J0003-2822 is normal for its spectral type.  However the Hipparcos distance would indicate that its absolute magnitude is overluminous by a factor of 1.5 for an M8.  This indicates, as noted in \citet{2007AJ....133..439C}, that 2MASS J0003-2822 is a potential near-equal luminosity unresolved binary which might affect the activity and age calculated from the \citet{West2009} relation (c.f. \citealt{2006AJ....131.1674S}). 

Based on the consistent age diagnostics of the primary and the secondary, an age of 0.9-1.4 Gyr is adopted for the system.  






\subsubsection{G 171-58 with 2MASS J00250365+4759191AB}
G 171-58 is an F8 star and lies 3.6 arcminutes from the L4+L4 close (separation 0.33\arcsec or $\sim$10 AU) binary dwarf 2MASS J0025+4759.  The possibility for companionship with G 171-58 was noted by  \citet{2006AJ....132..891R} and \citet{2007AJ....133..439C}.  G 171-58 is itself a spectroscopic binary (\citealt{2002AJ....124.1144L})  resolved in Hipparcos images with a separation of  $\sim$ 200 mas and an orbital period of just under 1 yr.  \citet{2008yCat.5128....0H}  measure [Fe/H]=0.22, and their age--metallicity relation suggests an age $<$ 2 Gyr. In this same study,  an age of 0.2 Gyr with an upper limit of 1.5 Gyr was estimated based on theoretical isochrones calculated from the T$_{eff}$, M$_{v}$, and [Fe/H] values.   The U, V velocities for G 171-58 fall into the Eggen box which also indicate an age $<$ 2 Gyr.  

The echelle spectrum of G171-58 shows clear K$_2$ maxima surrounding a central absorption core but the fairly low S/N coupled with the large magnitude of the photospheric contribution between the K$_1$ minima makes a direct measurement of R$^{\prime}_{HK}$ problematic. Instead, we undertook a differential analysis with respect to the F8 standard HD 187691, which has a measured log(R$^{\prime}_{HK}$)=-5.05 (Mamajek \& Hillenbrand 2008). We normalized the spectra in the Ca~II line wings and subtracted the spectrum of the standard. We convert the excess emission, seen in both the H and K lines, to surface flux, and add to this the log(R$^{\prime}_{HK}$)=-5.05 we had subtracted. We find that log(R$^{\prime}_{HK}$)=-4.81$^{+.03}_{-.08}$ for G~171-58, corresponding to $\tau_{2}$=2.2$^{+1.3}_{-0.4}$ Gyr.

\begin{figure*}[htbp]
\begin{centering}
\includegraphics[width=1.0\hsize]{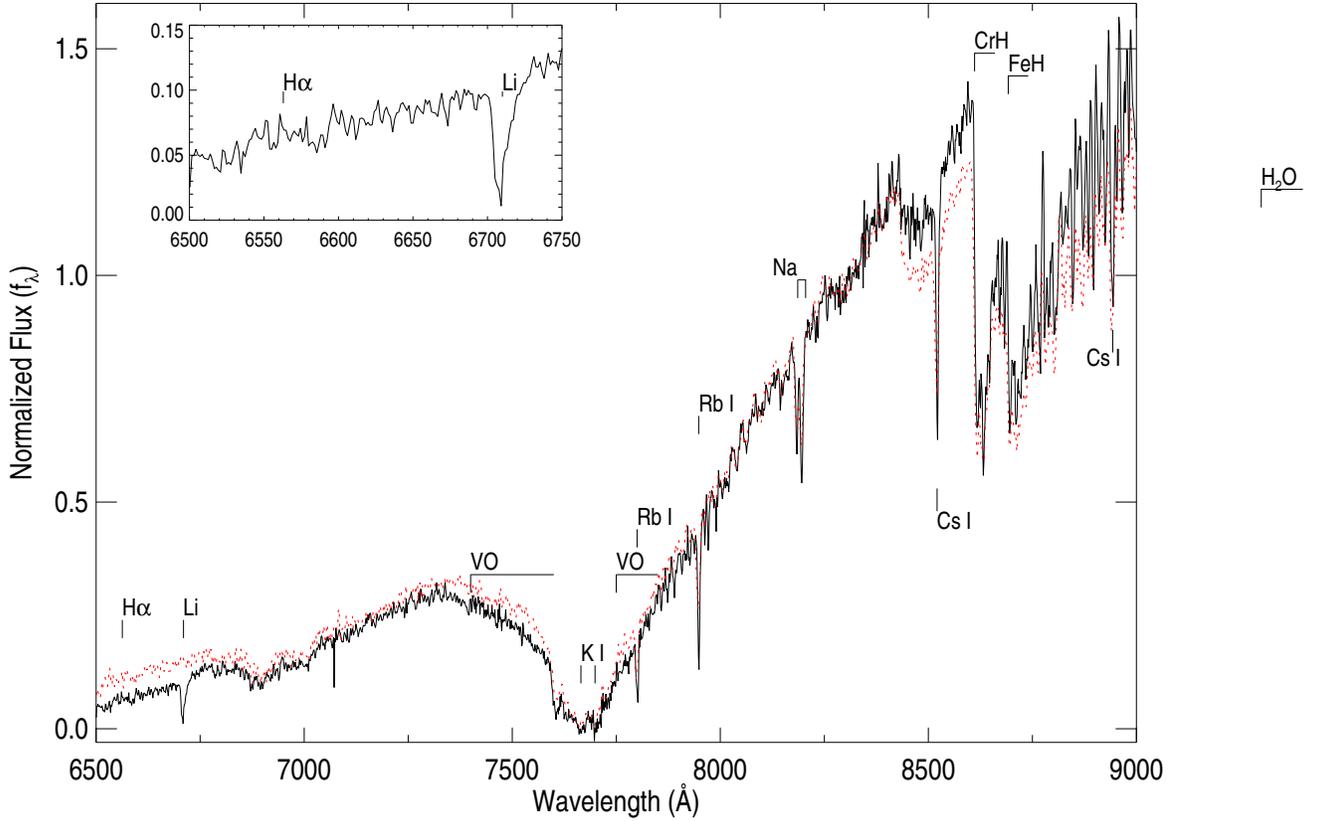}
\end{centering}
\caption{The optical spectrum of the secondary 2MASS J0025+4759 using published CTIO 4m data taken 2003 April 23 (\citealt{2007AJ....133..439C}).  The inset shows a lack of  H$\alpha$ (6563~\AA) emission but strong Li (6708~\AA) absorption.  As a reference, the LRIS optical spectrum of the standard L4 2MASSW J1155+2307 from \citet{2000AJ....120..447K} is over-plotted and normalized between 8240-8260~\AA~(dotted line).} 
\label{fig:2M0025+4759}
\end{figure*}

2MASS J0025+4759 is resolved into two near-equal mass components by \citet{2006AJ....132..891R}.  The combined spectrum exhibits Lithium absorption with an equivalent width of 10$\pm$2~\AA~ (\citealt{2007AJ....133..439C}) as seen in Figure~\ref {fig:2M0025+4759}, indicating component masses of at most $\sim$0.06M$_{\sun}$.  For an L4 spectral type at the bolometric luminosity calculated from the Hipparcos distance (see Table~\ref{Secondaries}) , this lead to an age upper limit of $\sim$0.5 Gyr for 2MASS J0025+4759 based on the evolutionary models of \citet{1997ASPC..119....9B}.  The J-$K_{s}$ color for 2MASS J0025+4759 is normal for its spectral type.  Despite the presence of Li absorption, the spectrum for this L4 companion does not display any low surface gravity features.  Therefore the age of this secondary is consistent with the range of 0.1-0.5 Gyr which is somewhat younger than indicated by the chromospheric activity of the primary.

We find a significant discrepancy between the age of the primary and secondary in this system.  2MASS J0025+4759 is likely younger than 0.5 Gyr and G 171-58 is likely older than 1.8 Gyr therefore we cannot adopt a suitable system age.  Rather we note the inconsistency in age diagnostics and calculate a mass for 2MASS J0025+4759 from the best age range of both the primary and the secondary.



\subsubsection{G 62-33 with 2MASS J13204427+0409045}
G~62-33 is a K2 dwarf based on the MagE spectrum.   The absence of Li absorption (W$_{\lambda}$(Li) $<$ 4 m\AA, logN(Li)$<$-2.9) in the optical spectrum indicates that this object is older than $\sim$ 1 Gyr.  The $\it{U,V}$ velocities fall outside of the Eggen box indicating an age $>$2 Gyr.  The metallicity for G 62-33 provides an upper bound on the age.  \citet{2008yCat.5128....0H}  determine [Fe/H]=0.15,  and \citet{2002yCat..33940927I} determine [Fe/H]=-0.18.   The majority of stars on the age-metallicity relation in \citet{2004A&A...418..989N} that lie between these two values are younger than 6 Gyr. 

The photometric data for G~62-33 shows that the star is clearly variable with peak-to-peak amplitudes increasing from 0.2~mag at $I$ to 0.4~mag at $B$.  However, we were unable to recover a unique period from the data which would have provided a gyro age.  A characteristic period from minimum to minimum is about 5 days for the first month of data, but this decreases to about 2-3 days during the last month. The  changes in variability amplitude with wavelength are consistent with a spotted surface and the apparent period change may be due to a rapid evolution of the spot structures.

We calculated log(R$^{\prime}_{HK}$) = -4.77$^{+0.05}_{-0.07}$ from the echelle data, where the uncertainties are dominated by uncertainties in the positions of the minima.  Comparison of the R-C data with three other K2 dwarfs observed at the same resolution with the R-C spectrograph, HD~22049, HD~4628, and HD~144628, independently showed that the emission strength lies between those of HD~22049 (log(R$^{\prime}_{HK}$)=-4.51, $\tau_2$=0.8~Gyr) and HD~4628 (log(R$^{\prime}_{HK}$)=-4.87, $\tau_2$=5.4~Gyr).   Although the 0.94 $B-V$ color of G~62-33 is slightly outside the quoted $B-V$=0.92 limit for the Mamajek \& Hillenbrand (2008) chromospheric/age relation, an extrapolation yields $\tau_2$=4.2$\pm$0.9~Gyr, consistent with the other age diagnostics. 

The L3 companion 2MASS J1320+0409 lies 1.1 arcminutes away from the primary.   The spectrum used to type this UCD has a very low signal to noise and leads to a $\pm$2 spectral type uncertainty.  However, unless this is an unresolved binary, the absolute magnitude calculated from the Hipparcos measurement is consistent with an L3 dwarf.  This object has a normal $J-K_{s}$ color for an L3.  It is difficult to ascertain whether the spectrum demonstrates low surface gravity features, or H$\alpha$ due to the low S/N.  Hence, no firm constraint of the age of the secondary can be made, but its photometric color suggests a middle-aged dwarf.

Since the age of the secondary is unconstrained, we adopt a system age of 3.3-5.1 Gyr based on the chromospheric activity of the primary.

\subsubsection[G 63-23 with 2MASS J13204159+0957506]{G 63-23 with 2MASS J13204159+0957506}

Based on a MagE spectrum, we classify G~63-23 as a K5 dwarf.  We place a 2$\sigma$ limit on the Li absorption equivalent width in our echelle spectrum  of $<$ 6~m\AA~which corresponds to log~N(Li)$<$-0.12 and a lower limit for the age of $\sim$1~Gyr.  There is no metallicity measurement to aid in constraining the age but the $\it{U,V}$ velocities fall outside of the Eggen box indicating an age $>$2 Gyr.   The photometric data for G~62-33 shows no significant periodic or quasi-periodic variability therefore gyrochronology can not be used.

From the echelle data of G~63-23, we determined  log (R$^{\prime}_{HK}$)= -4.49$^{+0.02}_{-0.03}$. At this spectral type, it is probably not wise to extrapolate the Mamajek $\&$ Hillenbrand (2008) age relation. Rather, we bound the age by comparing the activity level of G~63-23 with the K5 dwarfs $\xi$~Boo~B and 61~Cyg~A. \citet{2007ApJ...669.1167B} find gyrochronology ages for these two systems of $\sim$0.2 Gyr and $\sim$2 Gyr respectively and G 63-23 shows chromospheric activity between them albeit much closer to the level of the older star 61~Cyg~A.  Using the coefficients in Mamajek $\&$ Hillenbrand (2008) revises the age of 61~Cyg~A to 4 Gyr.  Assuming a Skumanich (1972)-like power-law decay of activity between 0.3 and 4 Gyr,
we find a likely age of G~63-23 of 1.2$\pm$0.4~Gyr.  Therefore we conservatively date this system as 1-3 Gyr which is roughly consistent with the Li and kinematic indications.

\begin{figure*}[htbp]
\begin{centering}
\includegraphics[width=1.0\hsize]{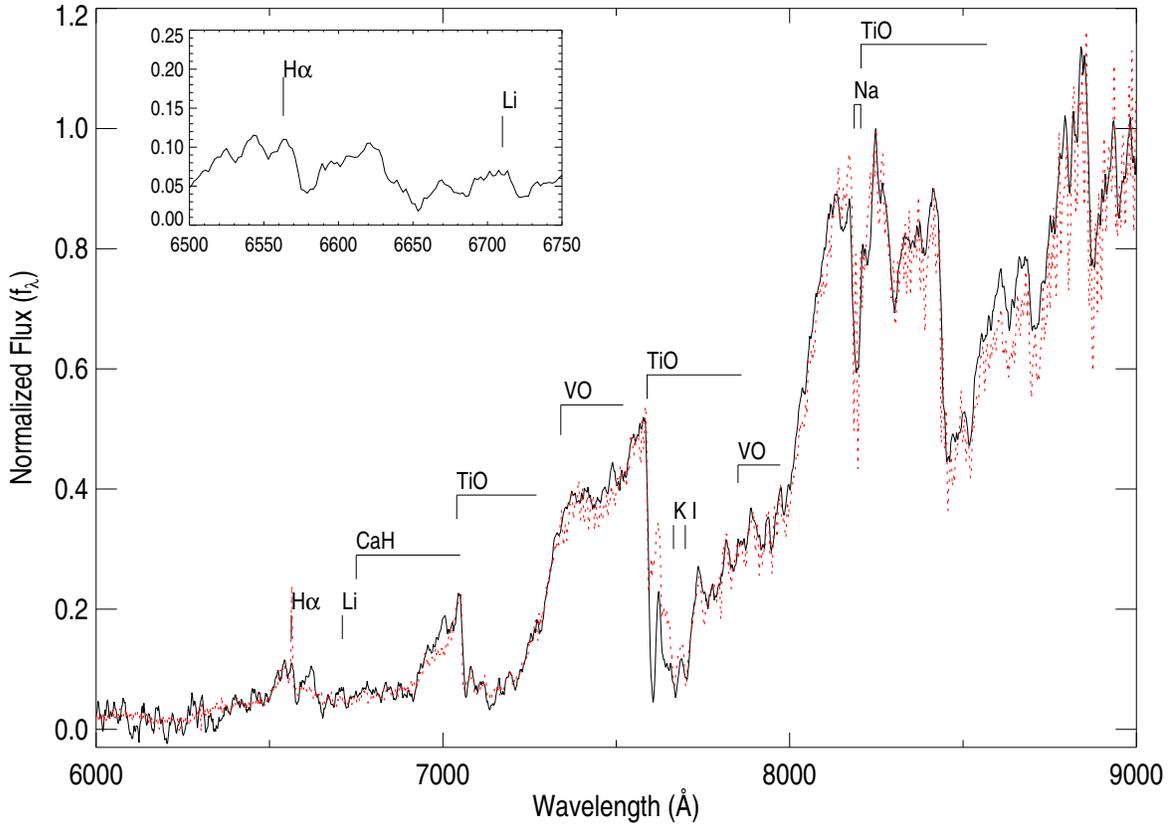}
\end{centering}
\caption{ The optical spectrum of the secondary 2MASS J1320+0957 using published CTIO 4m data taken 2003 April 20  (\citealt{2007AJ....133..439C}).  Over-plotted is the template for an M8 from \citet{2007AJ....133..531B} normalized at 8350~\AA~(dotted line).  The inset shows the region that contains H$\alpha$ (6563~\AA) emission and Li (6708~\AA) absorption neither of which are detected. } 
\label{fig:2M1320+0957}
\end{figure*}

2MASS J1320+0957 is an M8 dwarf that lies 2.8 arcminutes from G 63-23.  The J-$K_{s}$ color and v$_{tan}$ values for this object are normal for an M8.  We have re-examined a published spectrum from \citet{2003AJ....126.2421C} and find a lack of H$\alpha$ emission (W$_{\lambda}$(H$\alpha$) $<$ 300 m\AA) as seen in Figure~\ref {fig:2M1320+0957}. \citet{2008AJ....135..785W}  find that M7 dwarfs are active  for $8.0\pm^{0.5}_{1.0}$ Gyr. M dwarf activity increases with decreasing temperature through M7 dwarfs where, for the most part, all nearby objects show H$\alpha$ activity.  However, it is not clear that this trend continues at the cooler temperatures of M8 dwarfs and beyond where the photospheres become increasingly neutral (\citealt{2002ApJ...571..469M}; \citealt{2002ApJ...577..433G}).  So the lack of H$\alpha$ activity does not necessarily indicate that 2MASS J1320+0957 is  old for its spectral type.  As a result we can only assume a field M dwarf age range of 1-8 Gyr (\citealt{2009AJ....137....1F}) for this M8 dwarf. 

A system age of 1-3 Gyr is adopted for the G 63-23 and 2MASS J1320+0957 system based on the more reliable activity diagnostics of the primary.  However, while the kinematics and distance estimates for this system are in good agreement,  we are concerned of the age discrepancy between an H$\alpha$ inactive M dwarf and a chromospherically active K dwarf.

\subsubsection{G 200-28 with SDSS J141659.78+500626.4}
The primary in this system is a G5 star and it lies 9.5 arcminutes from the L5.5 dwarf SDSS J1416+5006.  \citet{2008yCat.5128....0H} determine a value for [Fe/H] of -0.16,  indicating a field age in the range of 1-5 Gyr.  They further determined an age range of 7-12 Gyr based on theoretical isochrones.  The kinematics of G 200-28 place this primary outside of the Eggen box for the young thin disk, in agreement with an age $>$2 Gyr.  There is no available measurement of Li absorption for this primary to aid in the age diagnosis.  

We obtained an echelle spectrum of the Ca~II H\&K region but despite fairly good S/N, the Ca~II emission cores are not clearly evident. The Ca~II line profiles are similar to those of the twilight sky, with deep central reversals. We place a limit of R$^{\prime}_{HK}<-5.0$, suggesting $\tau_2>6$~Gyr. G~200-28 appears older than the Sun. 

Therefore, based on the available diagnostics we adopt the theoretical isochrone estimated age range for G 200-28 of 7-12 Gyr.

SDSS J1416+5006 is classified as an L5.5 dwarf  by \citet{2006AJ....131.2722C}.  It has a spectral-type uncertainty of $\pm$2 based on a low signal to noise SpeX prism spectrum.  We have reanalyzed these data and deduce that an L4+/-1 is more likely.  The $J-K_{s}$ color of 1.56$\pm$0.09  for SDSS J1416+5006 is 0.18 magnitudes bluer than a normal L4 or L5 dwarf (\citealt{2009AJ....137....1F}) although the near-IR spectrum appears normal.  The blue near-IR color for its spectral type would indicate that SDSS J1416+5006 is likely to be older than the average UCD field population ($>$ 5 Gyr) or it is metal poor. However, the photometric uncertainty of this color does not allow a conclusive age constraint.

A system age of 7-12 Gyr is adopted for this system  from the isochrone analysis of the primary. 


\subsubsection{G 204-39 with SDSS J175805.46+463311.9}
The primary of this system is an M3 star that lies 3.3 arcminutes from the T6.5 dwarf SDSS~J1758+4633.  This primary is sufficiently late that solar-analog age/activity and age/rotation relations are not applicable, so we turn instead to the M dwarf age/activity relations examined by \citet{2008AJ....135..785W}.  \citet{2002AJ....123.3356G} measure an H$\alpha$ absorption equivalent width of -0.215~{\AA}.   Due to the cool atmospheres of M dwarfs, H$\alpha$ absorption is a sign of enhanced atmospheric heating and an indicator of magnetic activity.  However, the absorption phase likely represents the end of the active life of an M dwarf (\citealt{2009AJ....137.3297W}) indicating that G 204-39 is only weakly active.

It is also listed in the ROSAT All-Sky Faint Source Catalog (\citealt{2000yCat.9029....0V}) with a count rate of 2.53$\times$10$^{-2}$ cts s$^{-1}$, HR1=-0.58$\pm$0.18 and HR2=-1.0$\pm$0.27.  We used the  count rate/flux relation from \citet{1995ApJ...450..392S} to estimate the X-ray flux as 1.32$\times$10$^{-16}$ W m$^{-2}$.  The bolometric luminosity is calculated from the Hipparcos distance and combined with L$_{x}$ yields log(L$_{x}$/L$_{bol}$)=-5.3 which is slightly lower than the typical values for active M dwarfs (log(L$_{x}$/L$_{bol}$) $>$-4; \citealt{1995ApJ...450..401F}) .  Comparing to X-ray datasets of Hyades and Pleiades members where typical log(L$_{x}$/L$_{bol}$) values are $>$-4.5 for objects with similar colors,  G 204-39 appears to be older.  These measurements suggest that G 204-39 may be at the tail end of its active life, which \citet{2008AJ....135..785W} find to be 2.0$\pm$0.5~Gyr for M3 dwarfs. \cite{1990PASP..102..166E,1993AJ....106.1885E}  list G 204-39 as a member of the Hyades supercluster based on its proper motion and luminosity. Age estimates for this supercluster span a relatively broad range (e.g., \citealt{2001ASPC..228..398C} cite 0.5 to more than 2-3 Gyr) but since it is not a coeval sample (e.g. \citealt{2008A&A...483..453F,2007A&A...461..957F,2005A&A...430..165F}), it is not a useful age indicator.  The kinematics of G 204-39 do not indicate membership in the Hyades co-eval cluster and the chromospheric activity level discussed above further confirms that this object is likely older then $\sim$0.6Gyr.

The secondary of this system is the only T dwarf in our sample, and its properties have been studied in detail by \citet{2006ApJ...639.1095B} (hereafter BBK06) through a comparison of empirically-calibrated model spectral indices.  BBK06 find T$_{eff}$ = 960--1000~K and log g = 4.7--4.9 (cgs) for SDSS~J1758+4633, consistent with an age of 0.3-0.9 Gyr and at the low end of age estimates for the Hyades supercluster.  As companionship with a Hipparcos star provides a precise distance determination for SDSS~J1758+4633, we re-examined its properties as a check of the results of BBK06.  We first determined the luminosity of this source using the method described in \citet{2008ApJ...672.1159B}, by iteratively integrating its absolute flux-calibrated spectral energy distribution over the range 0.3--1000~$\micron$.  Near-infrared spectral data from BBK06 were used to calculate the 0.9--2.4~$\micron$ flux, after calibrating the data to $JHK$ photometry from \citet{2004AJ....127.3553K}.  The 2.4--9.3~$\micron$ flux was determined by piece-wise flux-calibrating a T$_{eff}$=1000~K, log g = 5.0 cgs spectral model from \citet{2005ApJ...624..988B} with mid-infrared photometry obtained with the {\em Spitzer Space Telescope} Infrared Array Camera (IRAC; \citealt{2004ApJS..154...10F}; program GTO-40198).  Apparent magnitudes of [3.6] = 14.88$\pm$0.04, [4.5] = 13.91$\pm$0.03, [5.8] = 13.64$\pm$0.10 and [8.0] = 13.15$\pm$0.04 were measured for SDSS~J1758+4633 from basic calibrated data (version S18.5.0) using IRAF PHOT and standard calibration methods for aperture photometry (\citealt{2005PASP..117..978R}).  Short- and long-wavelength fluxes were computed using a combination of spectral models and blackbody fluxes calibrated to the ends of the near-infrared and mid-infrared data.  This procedure provided a luminosity measurement of $\log(L_{bol}/L_{\sun})$ = -5.18$\pm$0.06, where the uncertainty includes astrometric and photometric uncertainties from the near-infrared and mid-infrared data, and systematic uncertainties in the luminosity calculation method (\citealt{2008ApJ...689L..53B}).

\begin{figure*}[htbp]
\begin{centering}
\includegraphics[width=1.0\hsize]{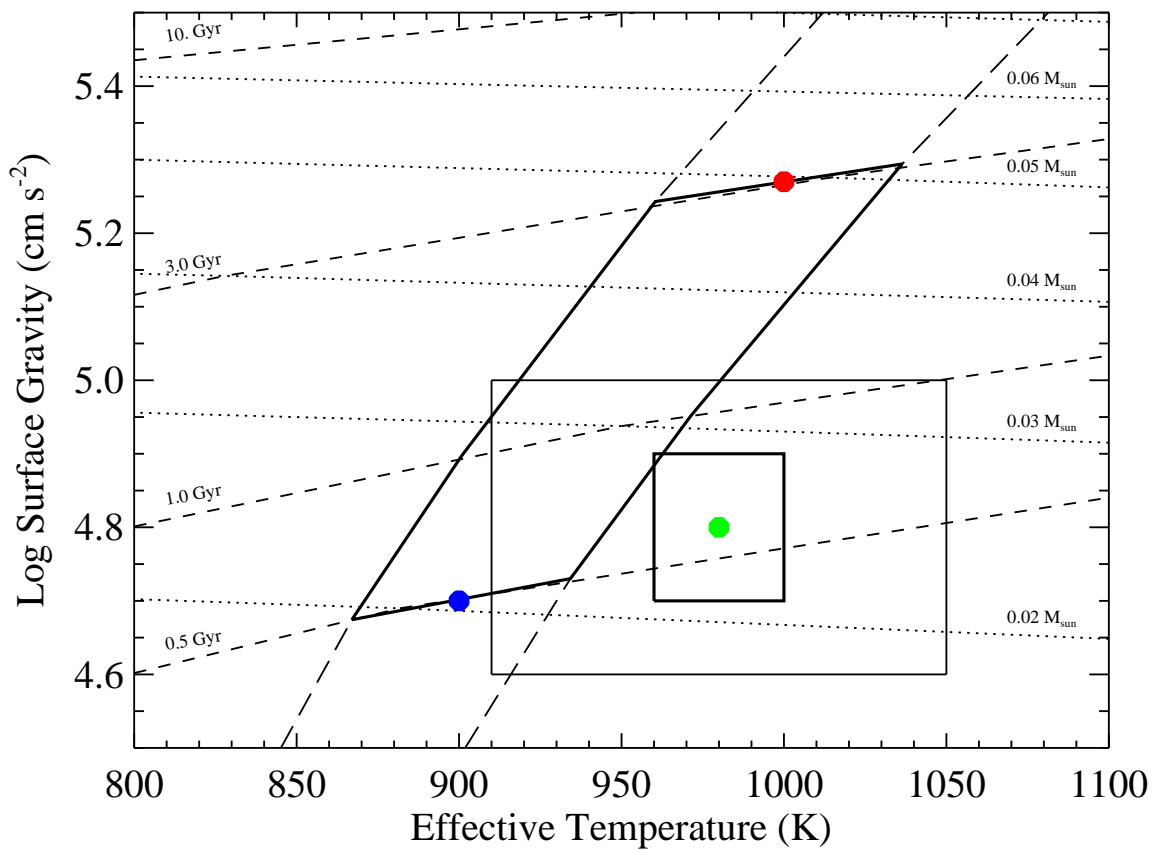}
\end{centering}
\caption{Effective temperature and surface gravity constraints for SDSS~J1758+4633, based on the measured luminosity of the source (vertical long dashed lines) and estimated age of the system (0.5--3~Gyr leftmost parallelogram), and based on the spectral index analysis of BBK06 (boxes to right; inner box indicates quoted parameter range, outer box indicates additional systematic uncertainties).  Labeled isochrone and isomass lines based on the evolutionary models of \citet{1997ASPC..119....9B} are indicated by short-dashed and dotted lines, respectively.  The green, red and blue circles correspond to spectral models shown in Figure~\ref{fig:2M1758}.
} 
\label{fig:tgphase}
\end{figure*}

\begin{figure*}[htbp]
\begin{centering}
\includegraphics[width=1.0\hsize]{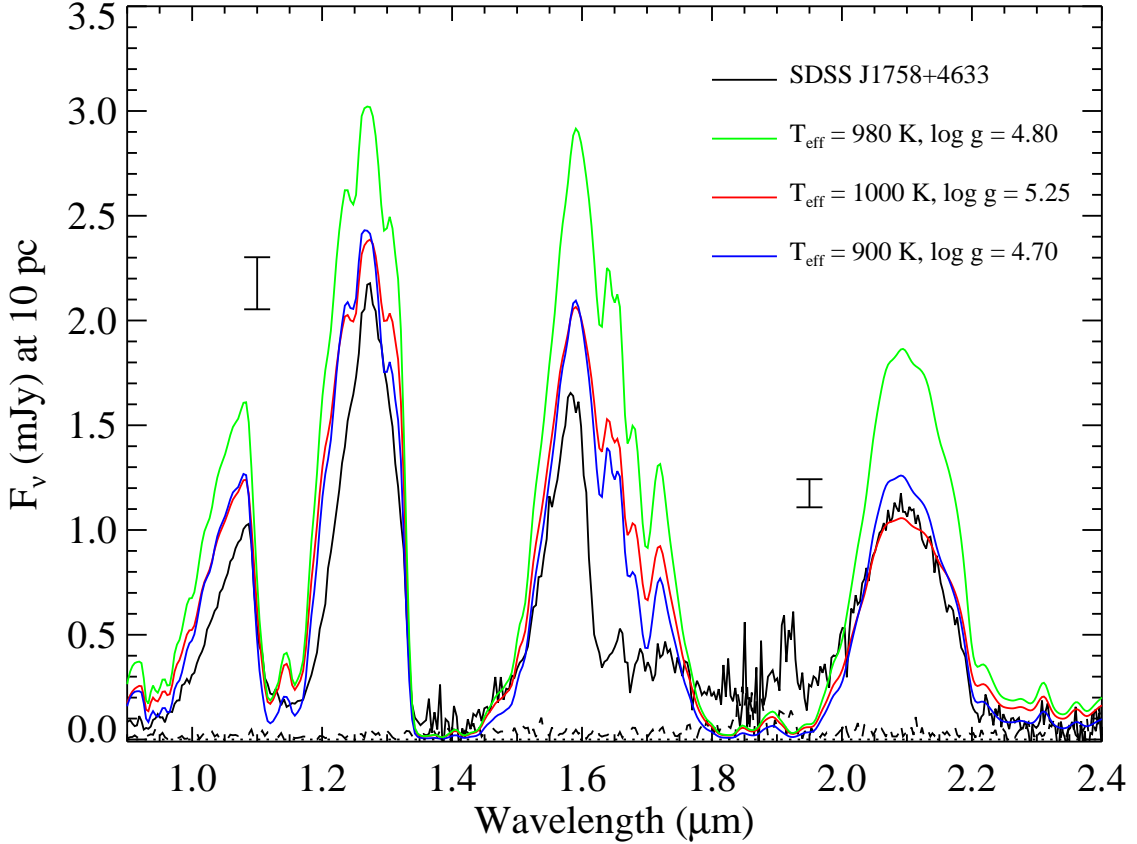}
\end{centering}
\caption{Comparison of observed absolute near-infrared spectral fluxes ($F_{\nu}$ at 10~pc) of SDSS~J1758+4633 (black line; dashed line shows uncertainties) to solar-metallicity spectral models from \citet{2005ApJ...624..988B} chosen from the T$_{eff}$/log g phase space constraint in Figure~\ref{fig:tgphase}.  The green line shows T$_{eff}$ = 900~K and log g = 4.80~cgs, based on spectral index constraints from BBK06; red and blue lines shows T$_{eff}$ = 1000~K and log g = 5.25~cgs
and T$_{eff}$ = 1000~K and log g = 4.70~cgs based on the luminosity/age constraints presented here.  All three models are scaled according to age- and mass-appropriate radii from \citet{1997ASPC..119....9B}.   Error bars indicate 1$\sigma$ uncertainties in the observed 1.27~$\micron$ and 2.1~$\micron$ flux peaks based on the absolute magnitudes of the source ($M_J$ = 15.20$\pm$0.06, $M_K$ = 15.46$\pm$0.06; HIPPARCOS, \citet{2004AJ....127.3553K}. 
Note that the two age/luminosity-constrained models have roughly the same scaling due to their common luminosity, while the BBK06-constrained model is too bright by $>$3$\sigma$.
} 
\label{fig:2M1758}
\end{figure*}

Combining just the luminosity measurement of the secondary, the age of the Hyades supercluster, and evolutionary models from \citet{1997ASPC..119....9B}, we derive an independent constraint on the T$_{eff}$ and log g of SDSS~J1758+4633 as shown in Figure~\ref{fig:tgphase}.  At the lower end of the age range, our analysis indicates T$_{eff}$ = 860--930~K, log g = 4.7~cgs and M = 0.02~M$_{\sun}$ for this source; at the upper end we find T$_{eff}$ = 910--1030~K, log g = 5.25~cgs and M = 0.05~M$_{\sun}$.  Note that the T$_{eff}$ estimates are broadly consistent with the $H-[4.6]$ = 2.29$\pm$0.04 color of this source (\citealt{2007MNRAS.381.1400W}).  Importantly, the T$_{eff}$/log g phase space constrained by the luminosity and age do not overlap with the seemingly tighter constraints provided by the BBK06 analysis.  Examination of the absolute spectral fluxes of SDSS~J1758+4633 appear to favor the luminosity analysis (Figure~\ref{fig:2M1758}); spectral models from \citet{2005ApJ...624..988B} tied to these constraints provide a closer match to the observed fluxes than models tied to the BBK06 constraints.  However, if the systematic uncertainties estimated for the BBK06 method are included ($\Delta$T$_{eff}$ = 50~K and $\Delta$log g = 0.1~cgs), there is reasonable overlap in T$_{eff}$ and log g constraints over the range 0.5--1.5~Gyr.   This somewhat younger age is consistent with enhanced $K$-band flux in the spectrum of SDSS~J1758+4633, indicative of reduced H$_2$ opacity (see BBK06).  However, it is also possible that this system is somewhat metal-rich, as indicated by comparison of CaH2+CaH3 and TiO5  for G 204-39 to other M3 dwarfs in \citet{2008AJ....135..785W} (G 204-39 has (CaH2+CaH3)/TiO5 of 2.45 where the range for M3 dwarfs was from 2.25-2.55).   Regardless, the activity level of the primary is consistent with the 0.5-1.5 Gyr age computed for the T dwarf, so we adopt this slightly younger age for the system.

\subsection{LSPM-N Pairs}
\subsubsection{NLTT 2274 with SDSS J004154.54+134135.5}
\begin{figure*}[!ht]
\begin{center}
\epsscale{1.5}
\plottwo{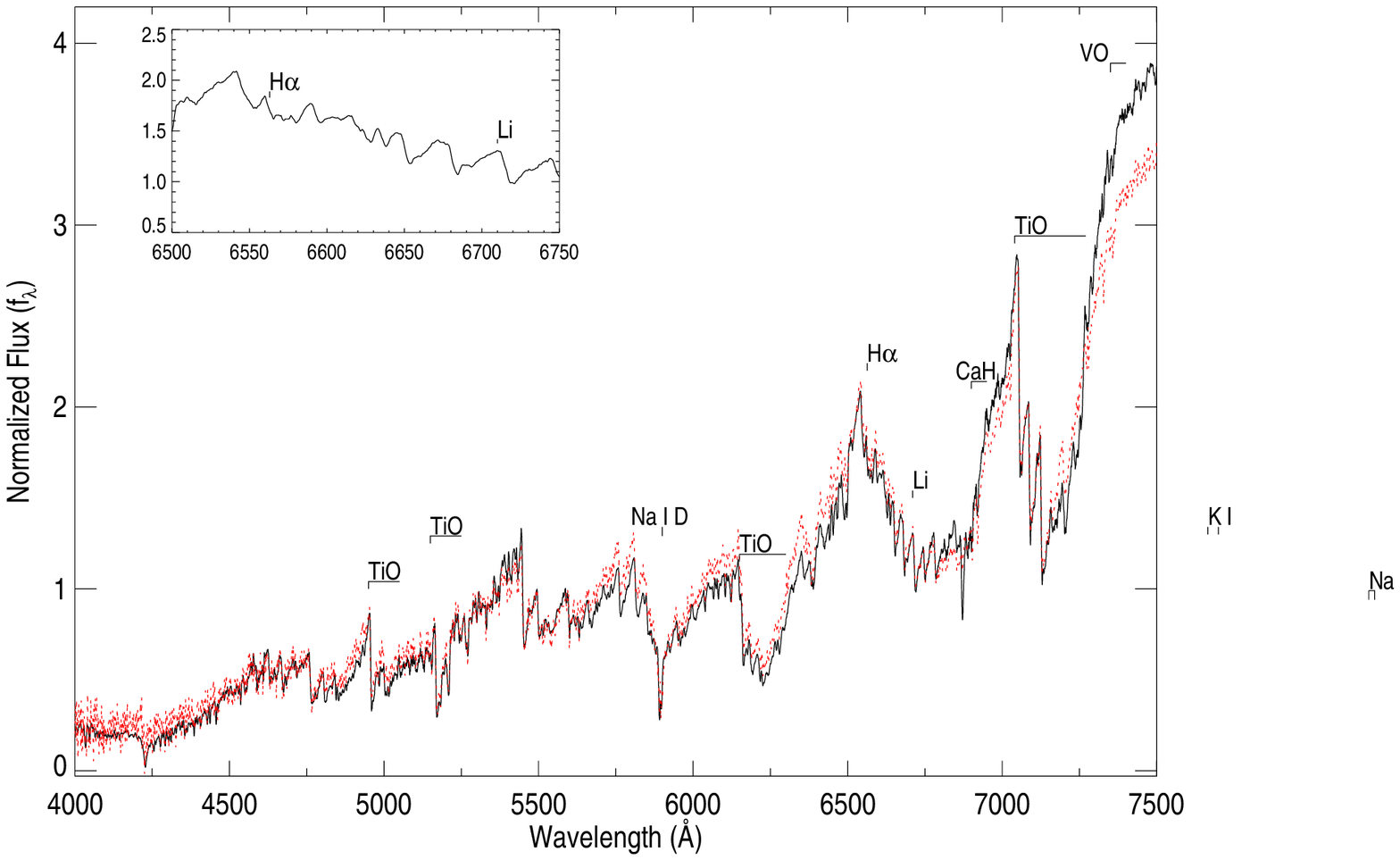}{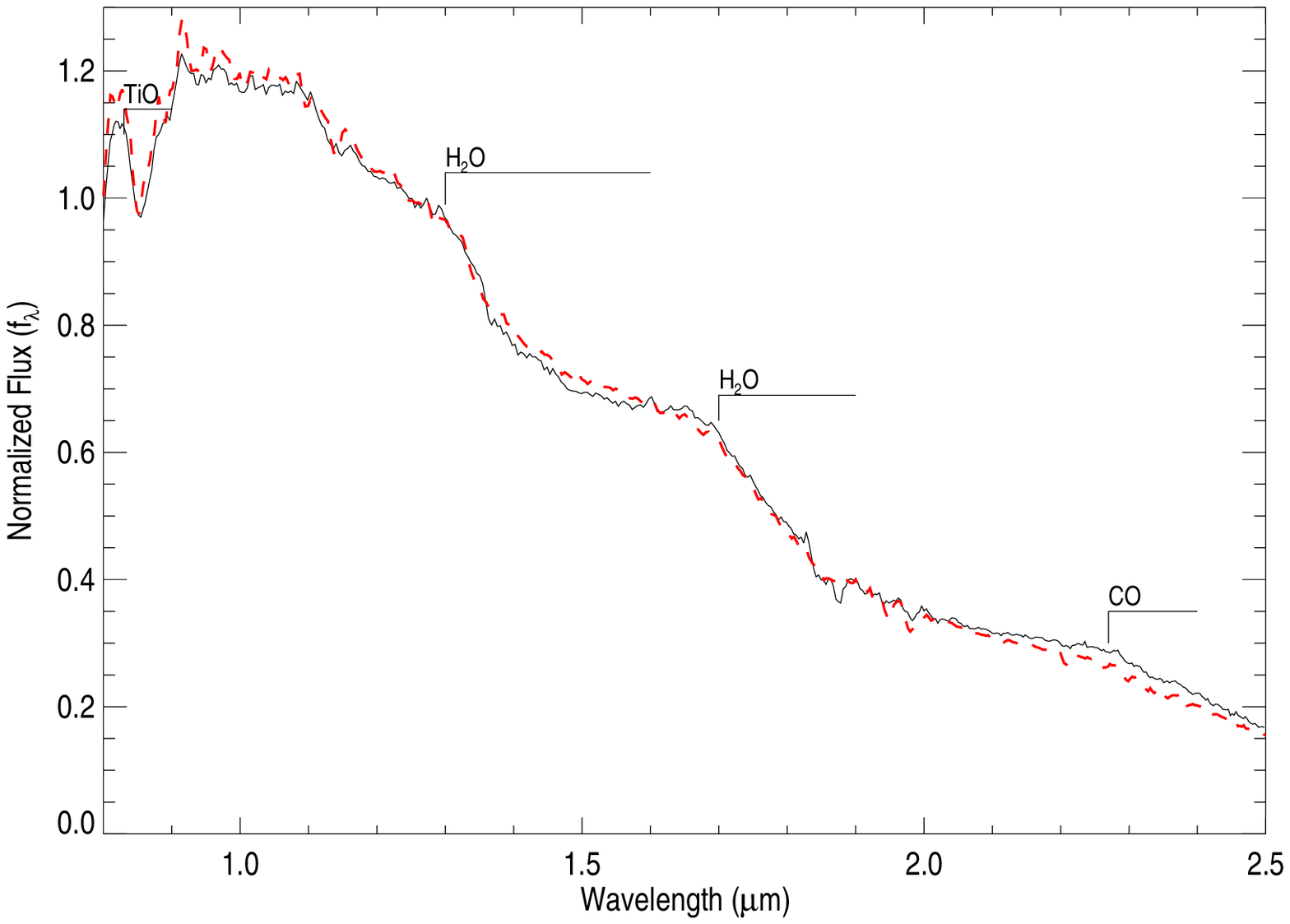}
\end{center}
\caption{The optical spectrum of the primary NLTT 2274 using MagE (top plot) and IR spectrum using SpeX (bottom plot).  Top: Over-plotted is the template for an inactive M4 from \citet{2007AJ....133..531B} (dotted line) normalized at 7400~\AA.  The inset shows a lack of both H$\alpha$ (6563~\AA) emission and Li (6708~\AA) absorption. Bottom:  Over-plotted is the M4 optical standard LP 508-14 (\citealt{2004AJ....127.2856B}) obtained from the SpeX Prism Library (dotted line).} 
\label{fig:2M0041+1341_Prim}
\end{figure*}

\citet{2008MNRAS.384.1399J} first noted this system as a potential wide pair due to its close separation (23$\arcsec$) and well matched proper motion components. Based on a MagE spectrum, we classify NLTT 2274 as an M4 dwarf (Figure~\ref{fig:2M0041+1341_Prim}). The 2MASS $\it{J}$ band relation from Golimowski et al. (2009) was used to calculate a spectro-photometric distance of  21$\pm$8 pc. This is in statistical agreement with the companion which has an estimated spectro-photometric distance of 31$\pm$6 pc. 

There is an absence of both Lithium absorption (W$_{\lambda}$(Li) $<$ 30 m\AA) and H$\alpha$ emission (W$_{\lambda}$(H$\alpha$) $<$ 100 m\AA) in the optical spectrum of NLTT 2274.  According to \citet{2008AJ....135..785W}, M4 objects remain active for 4.5$^{+0.5}_{-1.0}$Gyr so we use this as a lower bound on the age.  There is no radial velocity measurement available for this primary nor is there a defined metallicity relation for M dwarfs to further constrain the age.

\begin{figure*}[htbp]
\begin{centering}
\includegraphics[width=1.0\hsize]{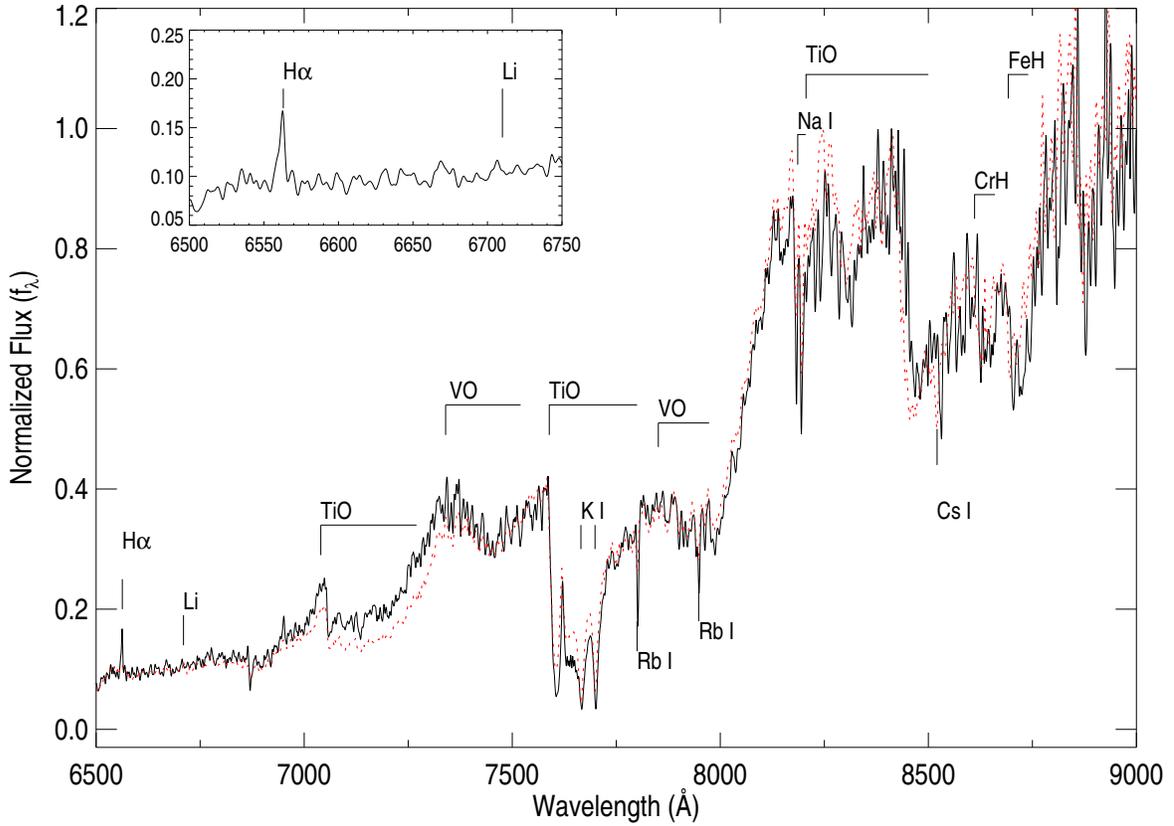}
\end{centering}
\caption{The optical spectrum of the secondary 2MASS J0041+1341 using MagE.  Over-plotted is the optical standard L0 dwarf 2MASP J0345+2540, from \citet{2000AJ....120..447K} with spectra normalized at 8350~\AA~(dotted line). The inset shows strong H$\alpha$ (6563~\AA) emission but no Li (6708~\AA) absorption. } 
\label{fig:2M0041+1341_Sec}
\end{figure*}

SDSS J0041+1341 was first identified in \citet{2002AJ....123.3409H} and classified as an L0 from a low signal to noise spectrum.  We re-observed this object with MagE and confirm the L0 spectral type (Figure~\ref{fig:2M0041+1341_Sec}).   H$\alpha$ emission was detected with an equivalent width of 2.2~\AA.  There is an absence of Li absorption (W$_{\lambda}$(Li) $<$ 400 m\AA) in the optical spectrum indicating a mass $>$ 0.06 M$_{\sun}$ and a corresponding age $>$ 0.5 Gyr.  SDSS J0041+1341 does not show any low-gravity features, such as weak Na or strong VO, indicating that it is older than 0.1 Gyr. The J-K$_{s}$ color and v$_{tan}$ values are both normal indicating it is a middle aged L dwarf (2-8 Gyr). 

The age-activity relations applicable to G and K dwarfs become more complicated in the late-type M and L dwarf regime.  As stars become fully convective ($\sim$ 0.35 M$_{\sun}$), the solar-type dynamo (\citealt{1993ApJ...408..707P,1955ApJ...122..293P}; \citealt{2003ARA&A..41..599T}) can no longer produce magnetic fields because the radiative-convective boundary (the tachocline) is not present to help generate and preserve the field.   However, the observed activity level of mid to late-type M dwarfs, which are beyond the fully convective boundary, remains high suggesting that a turbulent dynamo might be an alternate magnetic field source (\citealt{1993SoPh..145..207D}).  Indeed, recent MHD simulations have produced large-scale magnetic fields in fully convective stars (\citealt{2008ApJ...676.1262B}).  But while late-type M dwarfs are nearly all active, only a small fraction of L dwarfs have measured H$\alpha$ (\citealt{2004AJ....128..426W}); therefore these cooler objects mark a sharp change in activity.  \citet{2000AJ....120.1085G} and \citet{2007AJ....133.2258S} investigated whether active L dwarfs are likely to be younger than inactive L dwarfs at the same spectral type but their results were inconclusive.  It is likely that the drop in emission at the M/L transition is reflective of ineffective chromospheric heating as the photospheres become neutral (\citealt{2002ApJ...571..469M}; \citealt{2002ApJ...577..433G}; \citealt{2008ApJ...684.1390R}). This inactive M + active L system presents an interesting case for studying how the well-established age/activity relation for M dwarfs might break down at the cooler L dwarf temperatures.  Although we can not at this time rule out a binary interaction with an equal-magnitude or fainter companion as suggested for the active T dwarf  2MASSW J1237+6526 (\citealt{2000AJ....120..473B}),  SDSS J0041+1341 could demonstrate that an early type L dwarf can remain active at least through the activity lifetime of an M4 dwarf.  If the relationship between youth and H$\alpha$ emission breaks down for L dwarfs, activity metrics for these objects may prove to be poor indicators of age.

An age range of 4.5- 8 Gyr is adopted for the NLTT 2274 and SDSS J0041+1341 system based on the activity level of the primary and the upper age bound for normal field L dwarfs.

 \subsubsection{G 73-26 with SDSS J020735.60+135556.3}
\begin{figure*}[htbp]
\begin{centering}
\includegraphics[width=1.0\hsize]{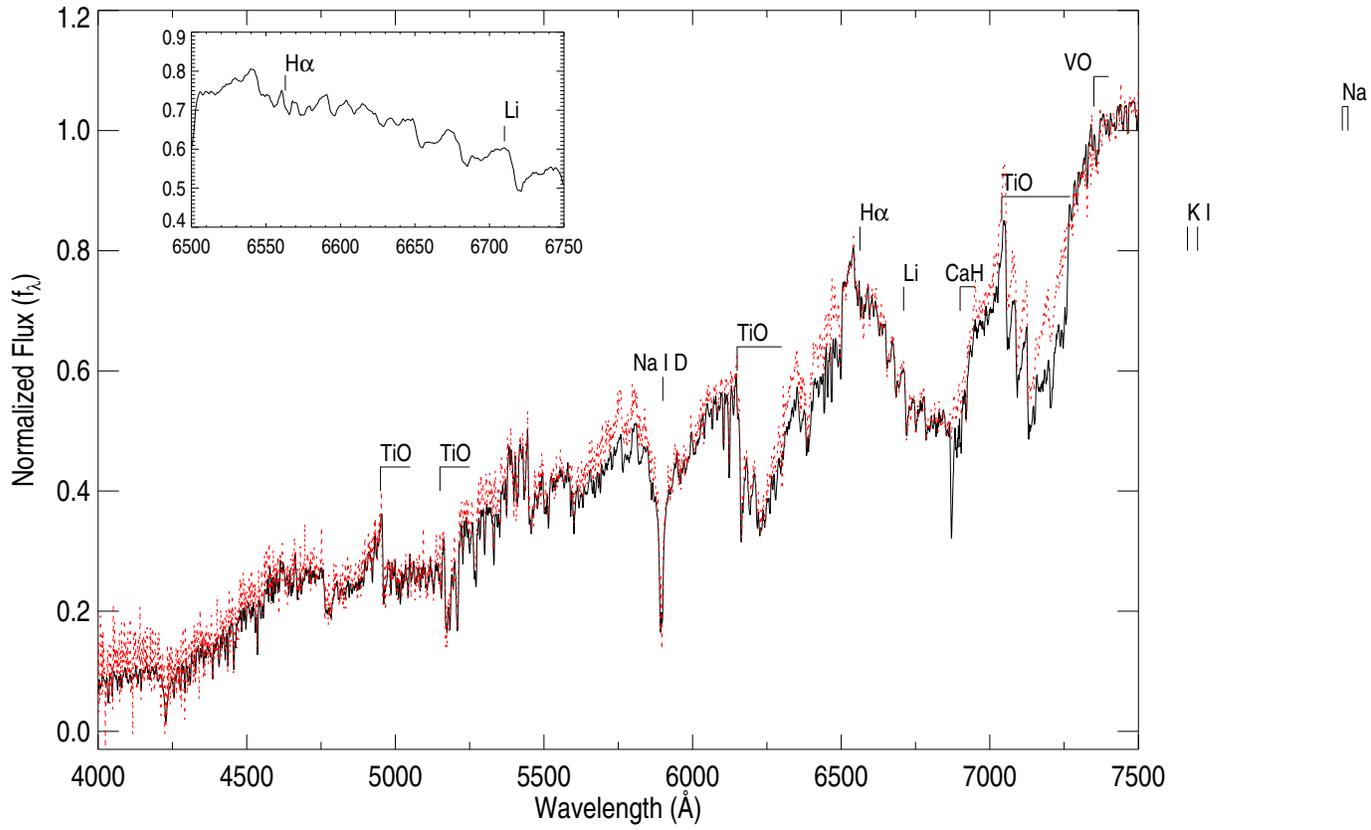}
\end{centering}
\caption{The optical spectrum of the primary LSPM J0207+1355 using MagE.  Over-plotted is the template for an inactive M2 from \citet{2007AJ....133..531B} normalized at 7400~\AA~(dotted line).  The inset shows a lack of both H$\alpha$ (6563~\AA) emission and Li (6708~\AA) absorption.} 
\label{fig:LSPM0207_Prim}
\end{figure*}

Based on our MagE spectrum, G 73-26 is an M2 dwarf (Figure~\ref{fig:LSPM0207_Prim}). The 2MASS $\it{J}$ band relation from Golimowski et al. (2009) yields a spectro-photometric distance of  26$\pm$10 pc. This is in statistical agreement with the L3 companion which has an estimated spectro-photometric distance of 35$\pm$5 pc.   There is an absence of both Li absorption (W$_{\lambda}$(Li) $<$ 40 m\AA) and H$\alpha$ emission (W$_{\lambda}$(H$\alpha$) $<$ -400 m\AA) in the optical spectrum. \citet{2008AJ....135..785W} determine that the active life of M2 stars ends at 1.2$\pm$0.4 Gyr placing a weak lower bound on the age.  A radial velocity (RV) of -107$\pm$13 km s$^{-1}$  was obtained for G 73-26 from an LDSS-3 spectrum.  Combining the photometric distance and available proper motion values with the RV yields ($\it{U,V,W}$)=(-44,-89,68) km s$^{-1}$ placing this object outside the Eggen box, favoring an age $>$ 2 Gyr.  

The $V$ and $I$ band modulation are small for G 73-26 and a shortest string analysis (\citealt{1983MNRAS.203..917D}) yields a likely period between 37 and 39 days. A sinusoidal fit to the $V$-band data yields a period of 39.6 $\pm$0.9~days with a semi-amplitude of 0.007$\pm$0.0007~mag and an $I$ band period of 39.6 $\pm$0.6~days with a semi-amplitude of 0.006$\pm$0.0003~mag.  The resultant gyro age is 3.4$\pm$0.5~Gyr, using the Mamajek \& Hillenbrand (2008) coefficients, which is consistent with an inactive M2 dwarf.

\begin{figure*}[!ht]
\begin{center}
\epsscale{1.5}
\plottwo{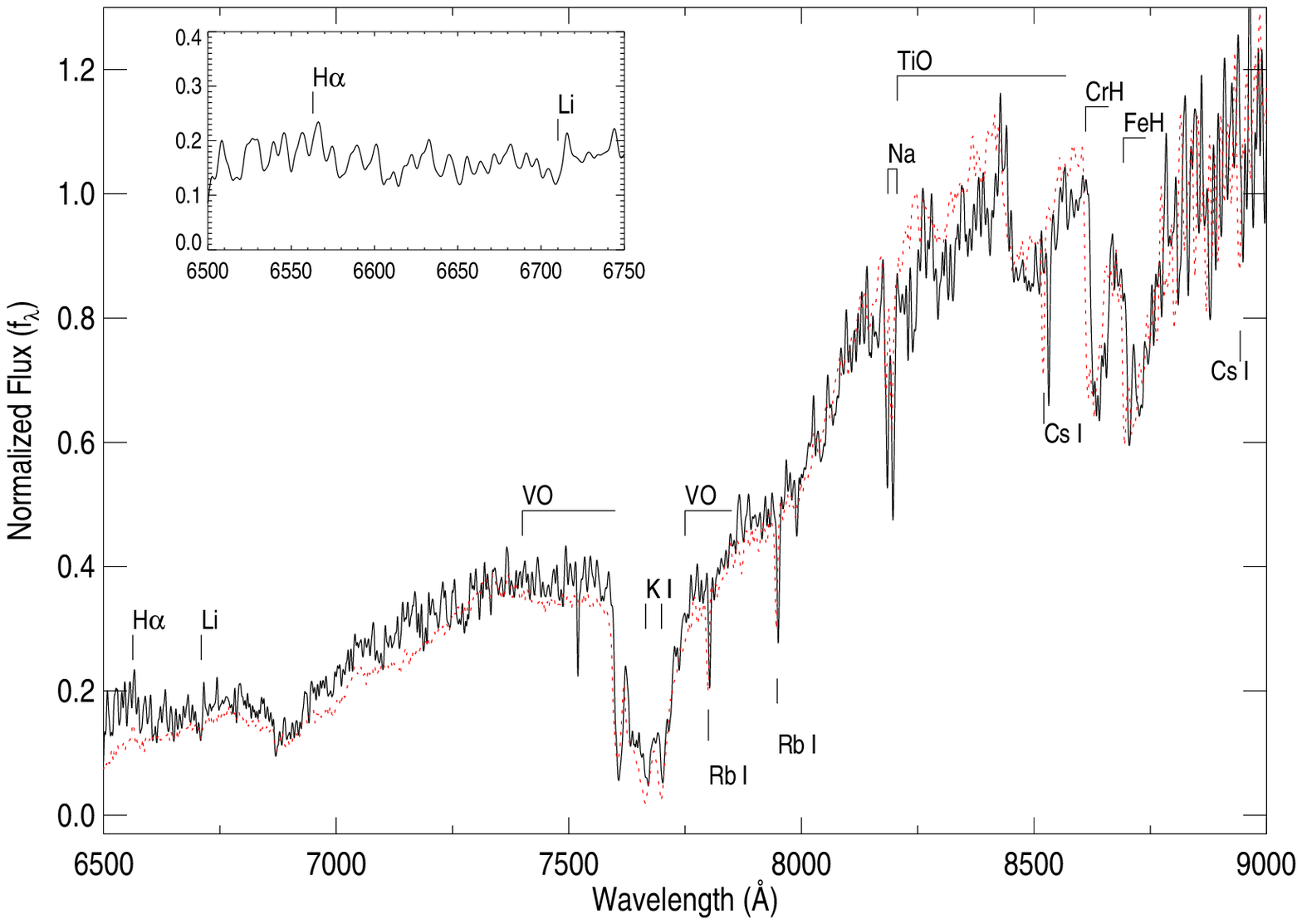}{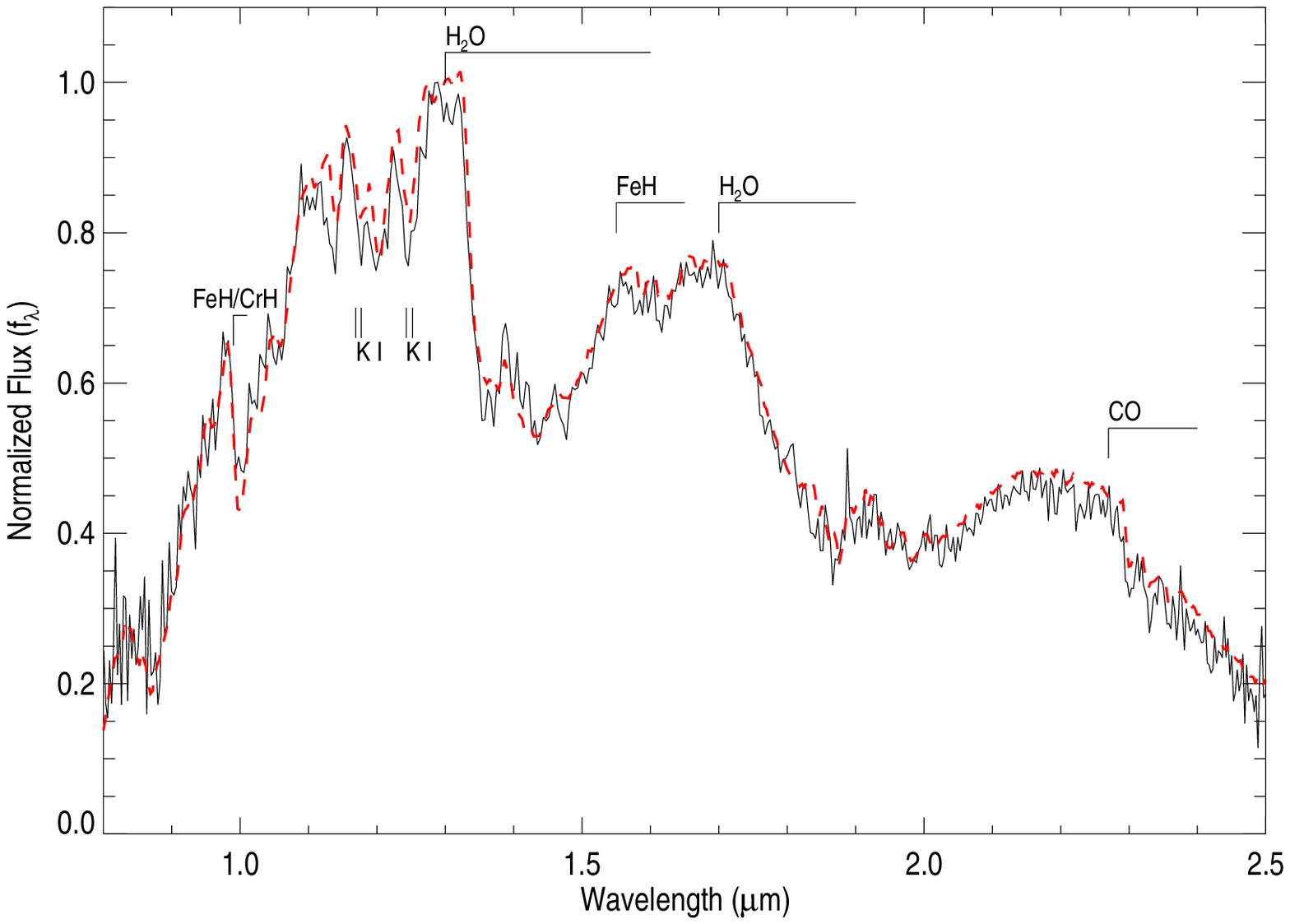}
\end{center}
\caption{The optical spectrum of the secondary 2MASSJ0207+1355 using MagE (top plot) and IR spectrum using SpeX (bottom plot).  Top: Over-plotted is Kelu-1, the L2 optical standard from \citet{1999ApJ...519..802K}, normalized between 8240-8260~\AA (dotted line).  The inset shows a lack of both H$\alpha$ (6563~\AA) emission and Li (6708~\AA) absorption. Bottom: Over-plotted is the L2 spectrum of SSSPM 0829-1309 (\citealt{2007ApJ...658..557B}) from the SpeX prism library (dotted line). } 
\label{fig:LSPM0207_Sec}
\end{figure*}

SDSS J0207+1355 was first identified as an L3 in \citet{2002AJ....123.3409H} and our MagE spectrum confirms this classification (Figure~\ref{fig:LSPM0207_Sec}).   There is an absence of both Li absorption (W$_{\lambda}$(Li) $<$ 200 m\AA) and H$\alpha$ emission (W$_{\lambda}$(H$\alpha$) $<$ 300 m\AA) in the optical spectrum. The J-K$_{s}$ color is normal for an L3 implying a field age in the range of 2-8 Gyr.

We adopt an age range of 3-4 Gyr for this system based on the rotation and activity level of the primary.

\subsubsection{G 121-42 with 2MASS J12003292+2048513}
From its optical spectra we infer that G121-42 is an M4 dwarf. There is a parallax measurement available which provides a distance of 32$^{+14}_{-7}$ pc (\citealt{1995gcts.book.....V}). The optical spectrum of G 121-42 lacks both Li absorption (W$_{\lambda}$(Li) $<$ 400 m\AA) and H$\alpha$ emission (W$_{\lambda}$(H$\alpha$) $<$ 200 m\AA).  \citet{2008AJ....135..785W} determine that the active life of M4 stars ends at 4.5$^{+0.5}_{-1.0}$Gyr placing a lower bound on the age of the system.   The photometric data shows clear long term sinusoidal variability in the $V$ band although no modulation is seen in the $I$ band.  The best fit sinusoid to the $V$ band data has a period of 47.0$\pm$0.9~days and a shortest string analysis (\citealt{1983MNRAS.203..917D}) shows a broad minimum at 46$\pm$3 days.  The semi-amplitude of the oscillation is 0.011$\pm$0.0007~mag.  The $B-V$ color of G121-42 is at the extreme of the stars Barnes (2007) used to derive gyro ages but still yields an age of 4.0$\pm$0.6~Gyr.

2MASS J1200+2048 is an active M7 with an H$\alpha$ equivalent width of 2.9~\AA~ (\citealt{2000AJ....120.1085G}; \citealt{2002AJ....123.2806R}).  We combine this value with the $\chi$ parameter from \citet{2004PASP..116.1105W} and measure log (L$_{H\alpha}$/L$_{Bol}$)=-4.8.   The age/activity relation of \citet{West2009} suggests an age range of 5-7 Gyr for this object.   \citet{2002AJ....123.2806R} found an absence of Lithium in the spectrum ($<$0.7~\AA), which is in agreement with an older field age. That study also calculated ($\it{U,V,W}$)=(-35$\pm$3,26$\pm$2,-32$\pm$1)  velocities for  2MASS J1200+2048 which  place it outside of the Eggen box favoring an age $>$ 2 Gyr.

Given these diagnostics we adopt an age for G 121-42 and 2MASS J1200+2048 of 4-5 Gyr.  This is slightly younger then the age predicted for 2MASS J1200+2048 from the H${\alpha}$ activity; however because the activity level of M dwarfs can be variable, this younger range is perfectly reasonable.


\subsection{UCD Mass Estimates}  
\begin{figure*}[htbp]
\begin{centering}
\includegraphics[width=1.0\hsize]{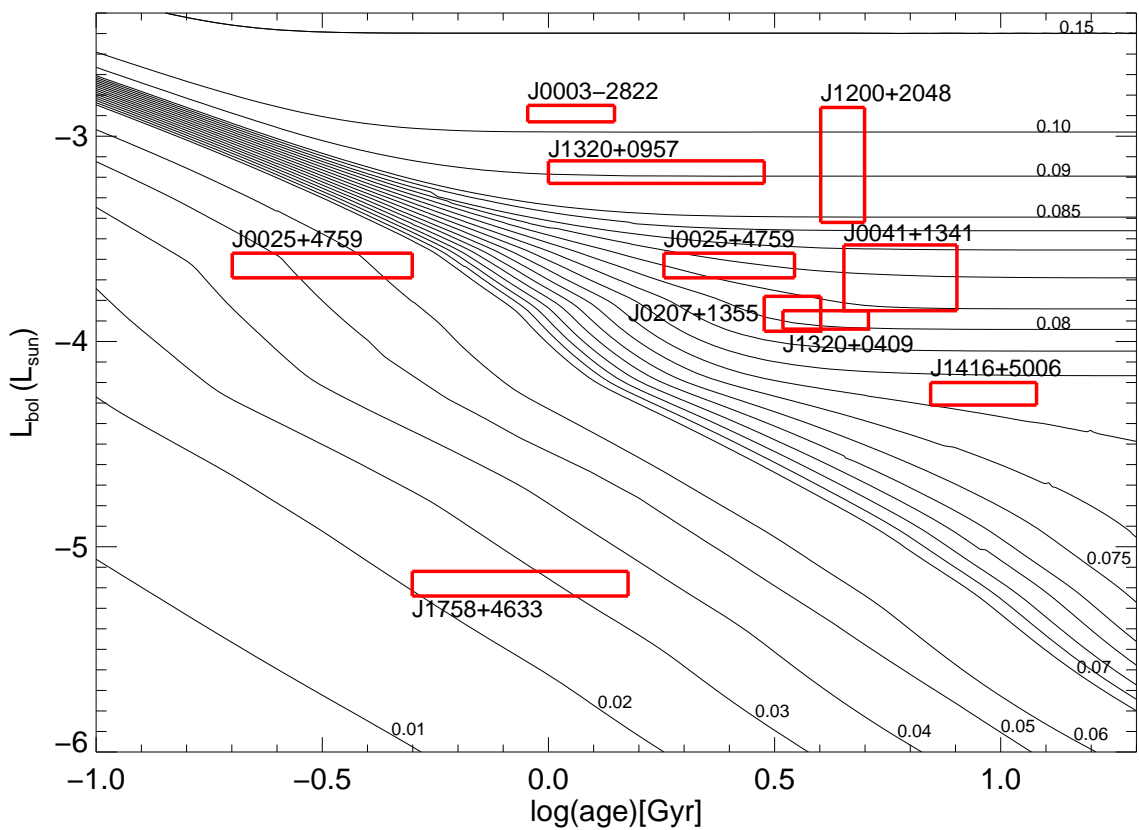}
\end{centering}
\caption{A plot of the \citet{1997ASPC..119....9B} evolutionary models with parameters (age and luminosity) for the nine candidate UCDs in wide pairs indicated with labeled boxes.  Masses from 0.01 through 0.15 M$_{\sun}$ are shown.   Only SDSS J1758+4633 is clearly of substellar mass. 2MASS J0025+4759 is listed twice due to the discrepancy in age diagnostics of the primary and the secondary in this system.  The box at left reflects the younger age calculated from the diagnostics of the secondary. The box at right reflects the older age calculated from the diagnostics of the primary.  } 
\label{fig:Burrows}
\end{figure*}

The evolutionary models from \citet{1997ASPC..119....9B} were used to estimate masses for the nine UCD secondaries.  Comparisons to the models were made using bolometric luminosities (L$_{bol}$), which were computed by combining distances (using parallax measurements or spectro-photometic distances) with apparent magnitudes and bolometric corrections with the exception of SDSS~J1758+4633, whose luminosity was calculated in Section 4.2.6.  For L and T dwarfs we converted K$_{s}$ apparent magnitudes from the 2MASS photometric system into the MKO system using the relations from \citet{2004PASP..116....9S}, and for M dwarfs we converted into the CIT photometric system using the color transformations from \citet{2001AJ....121.2851C}.  The bolometric corrections were calculated using either the relation from \citet{2004AJ....127.3516G} for L and T dwarfs or  from the measurements in \citet{2005nlds.book.....R} for M dwarfs.  Figure~\ref{fig:Burrows} shows the estimated age vs. L$_{bol}$  for the UCD secondaries against the evolutionary tracks from \citet{1997ASPC..119....9B}.  In general, masses for the substellar objects are very uncertain if the system age was poorly constrained due to the rapid change in brown dwarf luminosities with time.  We conclude that 2MASS J0003-2822, SDSS J0041+1341, SDSS J0207+1355,  2MASS J1200+2048, 2MASS J1320+0409, 2MASS J1320+0957, and SDSS J1416+5006 have masses above the hydrogen burning limit and are very low temperature stars at the bottom of the traditional stellar main sequence.  SDSS J1758+4633  falls below the hydrogen burning limit and is a brown dwarf.  2MASS J0025+4759 has a questionable age therefore an undetermined mass.  Table~\ref{Secondaries} lists our estimated ages, masses, and pertinent spectral characteristics for all of the UCD companions.




\section{DISCUSSION}
\subsection{Dynamic Stability and Maximum Separation Scales}
The separations of the nine companion systems discussed in this study are rather large for field UCDs and require a check as to whether or not they should have survived dynamical interactions within the Galaxy.  We investigated this question using the formalism of \citet{1987ApJ...312..367W} where the impact of perturbations from giant molecular clouds (GMCs) and close stellar encounters was examined for wide companion systems.  As in \citet{2003ApJ...586..512B}, and \citet{2003ApJ...587..407C, 2007ApJ...660.1492C}, the analytic solution of the Fokker Planck coefficients from \citet{1987ApJ...312..367W} describing the advective diffusion of a binary due to stellar encounters was used to investigate the sample. We work in the single kick limit\footnote{Assuming GM/$\epsilon$aV$^{2}_{rel}<<$(M/M$_{p}$)$^{2}$} and investigate the occurrence of disruptive encounters  using a rate which is proportional to mass and separation\footnote{In all calculations we use V$_{rel}$=20 km s$^{-1}$, $\epsilon$=0.1, n$_{\ast}$=0.1 pc$^{-3}$, n$_{GMC}$=4 x 10$^{-8}$ pc$^{-3}$,R$_{GMC}$=20 pc, M$_{GMC}$=5 x 10$^{5}$ M$_{\sun}$, N$_{clump}$=25 and M$_{p}$=0.7 M$_{\sun}$ as in \citet{1987ApJ...312..367W} and \citet{2007ApJ...660.1492C}  } as ${\rm f}_{cat}$ $\propto$ aM$^{-1}$.  All systems but that containing G 200-28 are subject to a frequency of disruptive encounters $<$ (20 Gyr) $^{-1}$.  G 200-28 has a frequency of $\sim$ (9  Gyr)$^{-1}$ which is approaching the inverse lifetime of the Galaxy and within our age range estimate for this system.  The characteristic diffusive timescale (t$_{\ast}\propto$ a$^{-1}$M) yields values $>$ 15 Gyr for all of the systems.  Therefore, close stellar encounters are not likely to affect these companions over the ages listed in Table~\ref{Ages}.  The impact parameter for interactions with giant molecular clouds is proportional to mass and separation as b$_{FP}^{GMC}$ $\propto$ M$^{-1/4}$a$^{3/4}$.  This value is larger than the maximum impact parameter b$_{max}$ $\propto$ a$^{3/2}$M$^{-1/2}$ for each of the nine companions, so such interactions are also not likely to have disrupted these systems. 

\begin{figure*}[htbp]
\begin{centering}
\includegraphics[width=1.0\hsize]{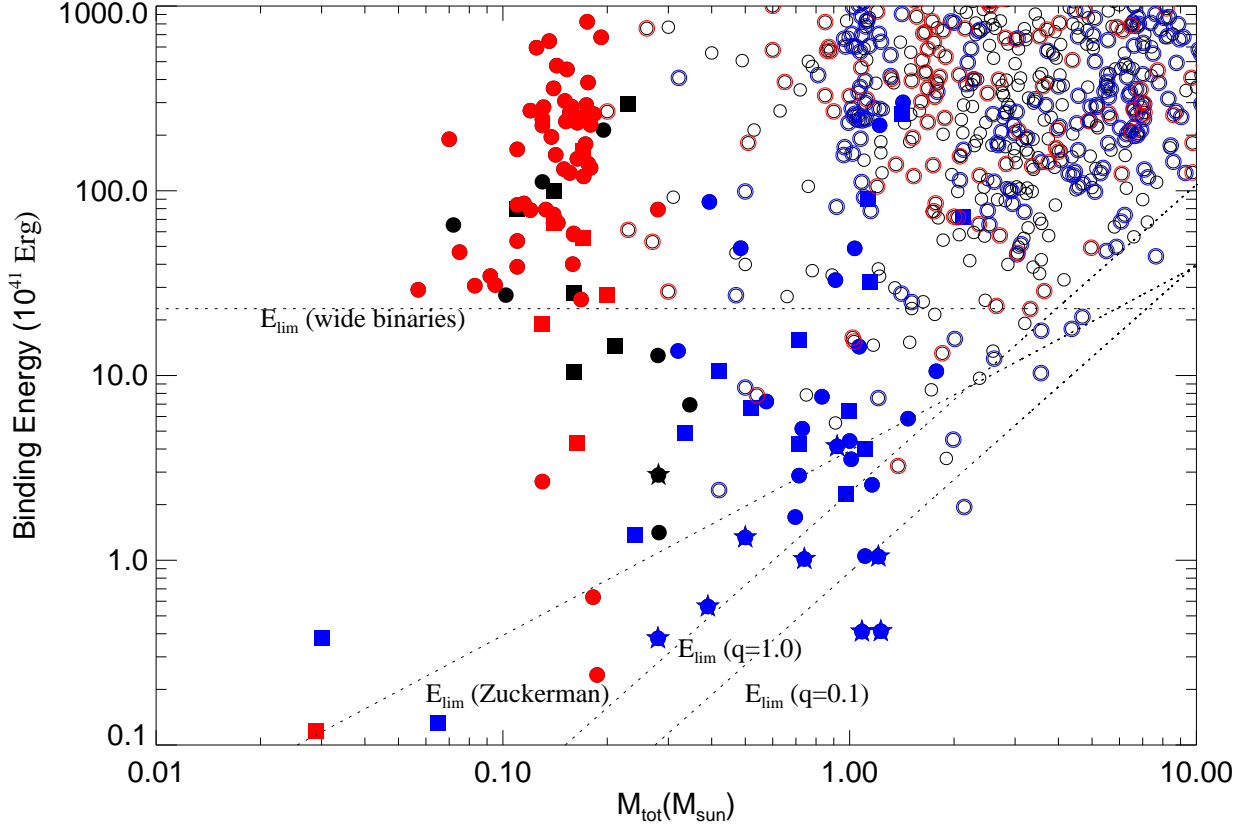}
\end{centering}
\caption{ System binding energy vs. total mass. Mass ratios are color coded on this plot with red symbols indicating q $>$ 0.7, blue symbols indicating q$<$0.3 and black symbols indicating 0.3$<$q$<$0.7.  Filled circles indicate systems containing a UCD, filled squares indicate systems containing a UCD that is younger than 500 Myr.  Open circles come from stellar companion catalogs.  The nine systems discussed in this paper are marked as five point stars. The minimum wide binding energy for brown dwarf field binaries (\citealt{2003ApJ...586..512B}; \citealt{2007ApJ...660.1492C}) is plotted as well as the minimum binding energy line from \citet{2009A&A...493.1149Z}. The two lines at the far right are our Jeans length criteria for q=1.0 and q=0.1 systems, respectively. } 
\label{fig:Binding_Energy2}
\end{figure*}

Recent results have shown that binding energies of the most weakly bound very low--mass (M$_{tot}<$0.2M$_{\sun}$) binaries in the field are $\sim$3 times larger than those of higher mass systems, suggesting a separation distribution of the field population that is sensitive to the conditions of formation. \citet{2003ApJ...586..512B}, and  \citet{2003ApJ...587..407C,2007ApJ...660.1492C}, find a minimum binding energy for very low--mass systems (nearly all of which have q $>$ 0.8 and M$_{tot}<$0.2M$_{\sun}$) of $\sim$2 x 10$^{42}$ ergs.   However this is clearly not the case for slightly more massive UCD systems.  Figure~\ref{fig:Binding_Energy2} shows the binding energy (E$_{b}$) versus total mass for a compilation of companion systems.  Stellar companions were gathered from the catalogs of \citet{1991A&A...248..485D}, \citet{1992ApJ...396..178F}, and \citet{1997A&AS..124...75T}; and young UCD companion systems from  \citet{2005ApJ...633..452K, 2006ApJ...649..306K}, \citet{2007ApJ...663..394K}, \citet{2009ApJ...691.1265L}, and \citet{2009arXiv0902.4742A}. Details on the field UCD systems were gathered from the Very Low Mass Binary Archive\footnote{http://vlmbinaries.org; see \citet{2007prpl.conf..427B} and references therein.}.   Table~\ref{All} lists the systems with widely-separated ($>$ 100 AU) UCD companions; i.e. those with the lowest binding energies.  The addition of recent systems both young and old with varying q values and small total mass complicates the idea of a minimum binding energy set at formation.   Four of the systems discussed in this study have 0.2M$_{\sun}<$M$_{tot}<$0.6M$_{\sun}$ but their binding energies are nearly ten times lower than the binding energies of the widest M$_{tot}<$0.2M$_{\sun}$ field systems.  Indeed, there are several field pairs now known with M$_{tot}>$0.1M$_{\sun}$ and E$_{b}$ $<<$ 2 x 10$^{42}$ ergs, as well as young, lower mass, weakly bound systems (e.g. \citealt{2007ApJ...660.1492C}; \citealt{2009A&A...493.1149Z}).  The system containing NLTT 2274 is especially interesting as it has M$_{tot} \sim$ 0.3M$_{\sun}$, E$_{b}<$ 10$^{42}$ ergs and an intermediate q value of $\sim$ 0.4.  These new systems indicate a gap in our sampling of intermediate mass companion systems, where a transition between weakly bound low--mass stellar companions and tight brown dwarf pairs occurs.

\citet{2009A&A...493.1149Z} have applied Jeans mass considerations to the problem of weakly bound very low- mass multiple systems.   Using the minimum fragmentation mass of a typical molecular cloud (7 M$_{Jup}$; \citealt{1976MNRAS.176..367L}) and assuming an arbitrary separation cutoff of 300AU, they derive a binding energy cut-off as shown in Figure~\ref{fig:Binding_Energy2}. However, a number of systems found in the field and young clusters violate this boundary indicating that 300 AU may not be a meaningful separation limit.   

Instead, we explored the Jeans mass criterion for wide companion systems using a Jeans length criterion to set the separation scale.  Two cases were examined: (1) q=1.0 with the maximum separation equal to twice the Jeans length; (2) q=0.1 with the maximum separation equal to the sum of the Jeans length for a system of mass  M$_{2}$ and a system of 10 x M$_{2}$.  We are assuming that the minimal initial separation of a pair that formed together should roughly equal the Jeans length.  Subsequent dynamics such as gravitational infall and scattering or sub-fragmentation at the time of formation will generally bring sources closer together and perturbations from Galactic encounters will generally pull systems further apart.  However the Jeans length is a good starting point for the widest separation of companions that formed from the same molecular cloud.  The resultant binding energy cut-offs are shown in  Figure~\ref{fig:Binding_Energy2}.  The difference between them is small and all but two of the systems discussed in this paper have binding energies that fall within the maximum scale set by the first fragmentation stage. Indeed the distribution of all systems shown in Figure ~\ref{fig:Binding_Energy2} are well-constrained by these lines over 0.2M$_{\sun}$$<$M$_{tot}<$10M$_{\sun}$, suggesting that this variable separation scale is a more realistic limit than an arbitrary fixed separation limit.  This envelope does not attempt to explain why field systems with M$_{tot}$ $<$ 0.2M$_{\sun}$ are almost all at significantly tighter separations than what is predicted by the Jeans criterion.   It may be that dynamical effects are more important in the initial formation of such low--mass objects than for more massive stellar systems (\citealt{2001AJ....122..432R}; \citealt{2002MNRAS.336..705B}), although we still cannot rule out insufficient sampling of the parameter space.

\subsection{Higher Order Multiplicity Among Wide Systems}
One explanation for the unusually low binding energies for some of the UCD systems plotted in Figure~\ref{fig:Binding_Energy2} is that one or both components may themselves be unresolved multiples.   It has been suggested by \citet{2005AJ....129.2849B} that there is a higher binary frequency among brown dwarfs when they are found widely-separated from a common motion stellar primary versus those found isolated in the field.  The larger binary fraction could be indicative of a formation mechanism which requires a higher order multiple system to keep all components gravitationally bound, or requires an exchange of angular momentum between wide and close components.  

We have re-visited this conjecture with the objects listed in Table~\ref{All}.  There are 44 systems containing a UCD which is over 100 AU from the primary star, of which 20 have had their UCD secondaries targeted with adaptive optics or the Hubble Space Telescope to search for additional components down to 0.1$\arcsec$-0.5$\arcsec$ separations.  This higher resolution probes the projected separation space within 20 AU, which characterizes the majority ($\sim$90\%) of UCD binaries (\citealt{2007ApJ...668..492A}). From this subset we calculate a resolved binary frequency for UCDs of  $\epsilon_{b}$=50$\pm11\%$.  This is significantly larger than the resolved binary frequency for field UCDs, which typically range over 10-20\% (e.g. \citealt{2001AJ....121..489R,2006AJ....132..891R}; \citealt{2003ApJ...586..512B, 2006ApJS..166..585B}; \citealt{2003AJ....126.1526B};  \citealt{2003ApJ...598.1265S,2007AJ....133.2320S}; \citealt{2003ApJ...587..407C}).  Our UCD multiplicity fraction is a $\sim$2$\sigma$ deviation from the field, consistent with the results of \citet{2005AJ....129.2849B}.  We note that our companion sample is not a volume-complete one and complex selection effects (other than those associated with formation mechanisms) may be present.  In the worst case scenario of a magnitude limited sample that favors unresolved multiples, the binary fraction of isolated field sources increases to 10-30\%. But this is still well below the $\sim$50\% binary fraction found for the wide multiples. 
This large fraction of triples is surprising.  

For comparison, the 8pc sample (\citealt{2005nlds.book.....R}) contains 118 M dwarfs, with 55 single stars, 26 binaries, 6 triples, and 1 quadruple system. Hence the ratio of triples to binaries is roughly 1:4 and quadruples to binaries is 1:26.  Our wide UCD companion sample includes 20 binaries, 12 triples, and 5 quadruples so we find these ratios to be 3:5 and 1:4 respectively.  The addition of a third or fourth component to these wide binaries may be required to maintain the stability of the system.  This high rate of multiplicity is also relevant to the binding energies plotted in Figure~\ref{fig:Binding_Energy2}, as the addition of an unseen UCD or stellar companion could increase binding energies by as much as 50\%.  

\subsection{ Discrepancy Among the Ages}

Establishing common proper motion, distance and radial velocity are important checks on the likelihood of a co-eval pair.   However, for the UCD population, precise distances are difficult to establish and radial velocities are rare.  Consequently, establishing a common age via activity, kinematic and/or metallicity diagnostics becomes a particularly important tool for confirming companionship.   But, as seen in this work, discrepancies still arise among the available age diagnostics.  While the differences in ages discussed in section 4 of this paper do not seem large enough to force us to disregard possible companions, they do serve as intriguing cases for examining current age-activity relations for both stellar and substellar objects.  For instance,  G 171-58 has chromospheric activity levels which likely place it as older then $\sim$1~Gyr while its companion, 2MASS J0025+4759, has strong Li absorption and is most likely younger then $\sim$0.5 Gyr.   G 63-23 has both chromospheric and rotation ages which suggest it is younger than $\sim$3 Gyr while its companion, an H$\alpha$ inactive M8, resembles an older field star.  NLTT 2274 is a mid-type M dwarf which shows no H$\alpha$ activity making it among the older field stars while its companion is an H$\alpha$ active L0.  These pairs may end up as excellent tests for the low mass-star and substellar activity relations.   Regardless, we encourage future investigations of UCD companions to carefully examine coevality of a proposed system before assuming companionship.

\section{CONCLUSIONS}
We have provided a detailed analysis of nine wide companion systems containing UCDs. Seven of the systems have parallax measurements, six of which are precise Hipparcos measurements.  Combining catalog information with new spectroscopic observations of the primary and secondary components,  a best age range for each system was determined.  Assuming co-evality with the secondaries and combining best age ranges with bolometric luminosity ranges, masses were estimated from the \citet{1997ASPC..119....9B} evolutionary models.  Seven of the UCDs were determined to be very low--mass stars, one was determined to be substellar, and one has a questionable age therefore undetermined mass.  Two of the nine systems, G171-58 with 2MASS J0025+4759 and G 63-23 with 2MASS J1320+0957,  have significant differences in the component system ages indicating possible shortcomings in our understanding of the age diagnostics of stars and ultracool dwarfs. 

Using a compiled list of known wide companion systems containing a UCD, we find that the frequency of tight resolved binaries is at least twice as high for wide companion UCDs as for isolated field equivalents.   The ratio of triples to binaries is 3:5 and quadruples to binaries is 1:4 for wide companion systems with resolved UCD secondaries versus 1:4 and 1:26 for the 8-parsec sample.  The higher frequency of higher order multiples suggests that a third or fourth component may be required to maintain gravitational stability or to facilitate the exchange of angular momentum in these loosely bound systems.

The Jeans criterion was investigated against a large sample of companion systems and we conclude that using the Jeans length to set the separation scale is sufficient for constraining the lowest binding energy UCD companion systems down to M$_{tot}\sim$0.2M$_{\sun}$.  However, the tight separation of the closely bound, near equal-mass UCD systems is not explained by the allowed envelope set by the Jeans length. The distinguishing characteristics of objects now known at varying mass ratios, total masses, separations, and ages suggests that more specific predictions from relevant theories will help distinguish the dominant formation mechanism for the UCD population.



	
\acknowledgments{}
	



\acknowledgments{The authors would like to thank S. Lepine and C. P. McNally for useful conversations about companion systems, E. Mamajek for a useful discussion on supercluster membership as well as a thorough and helpful referee report, and T. Dupuy for a helpful exchange on known UCD companions. Research has benefitted from the M, L, and T dwarf compendium housed at DwarfArchives.org and maintained by Chris Gelino, Davy Kirkpatrick, and Adam Burgasser;  this publication has made use of the VLM Binaries Archive maintained by Nick Siegler at http://www.vlmbinaries.org and the SPEX Prism Spectral Libraries, maintained by Adam Burgasser at http://www.browndwarfs.org/spexprism.  This publication makes use of data products from the Two Micron All-Sky Survey, which is a joint project of the University of Massachusetts and the Infrared Processing and Analysis Center/California Institute of Technology, funded by the National Aeronautics and Space Administration and the National Science Foundation. This research has made use of the NASA/ IPAC Infrared Science Archive, which is operated by the Jet Propulsion Laboratory, California Institute of Technology, under contract with the National Aeronautics and Space Administration. We acknowledge receipt of observation time through the SMARTS consortium.  We also would like to thank SMARTS queue observers M. Hernandez and J. Velazquez. }

\begin{deluxetable}{lllllllllllll}
\label{tab:tab1}
\tabletypesize{\tiny}
\rotate
\tablecaption{Astrometric Information on previously studied companion systems containing a UCD\label{All}}
\tablewidth{0pt}
\tablehead{
\colhead{Name} &
\colhead{SpT} &
\colhead{SpT} &
\colhead{$\rho_{star-UCD}$}  &
\colhead{$\rho_{star-UCD}$} &
\colhead{$\rho_{UCD-UCD}$\tablenotemark{a}} &
\colhead{$\rho_{UCD-UCD}$\tablenotemark{a}} &
\colhead{Lower Age\tablenotemark{b}} &
\colhead{Upper Age\tablenotemark{b}} &
\colhead{Mass ($M_{\sun}$)\tablenotemark{c}}&
\colhead{Mass ($M_{\sun}$)\tablenotemark{d}} &
\colhead{Ref } \\

& 
\colhead{(primary)} &
\colhead{(secondary)} &
\colhead{($\arcsec$)} &
\colhead{(AU)}&
\colhead{($\arcsec$)} &
\colhead{(AU)}&
\colhead{(Gyr)} &
\colhead{(Gyr)} &
\colhead{(primary)}&
\colhead{(secondary)} \\
\colhead{(1)} &
\colhead{(2)} &
\colhead{(3)} &
\colhead{(4)} &
\colhead{(5)} &
\colhead{(6)} &
\colhead{(7)} &
\colhead{(8)} &
\colhead{(9)} &
\colhead{(10)} &
\colhead{(11)} &
\colhead{(12)} \\
}
\startdata
TWA 5\tablenotemark{g}			&	M1.5			&	M8		&	2	&	100	&	$<$0.15	&	$<$8		&	0.01	&	0.3	&	0.40	&	0.02	&				    3\\
GQ Lup\tablenotemark{g}			&	K7			&	L1.5		&	0.7	&	103	&	$<$0.4	&	$<$71	&	3	&	3	&	0.70	&	0.02	&			20\\
G203-50\tablenotemark{g}		&	M4.5			&	L5		&	6.4	&	135	&	$<$0.18	&	$<$11	&	1	&	5	&	0.15	&	0.07	&			32\\
LHS 5166						&	dM4.5		&	L4		&	8.43	&	160	&	$<$1		&	$<$24	&	2.6	&	8	&	0.21	&	0.07	&			25\\
GJ 1001\tablenotemark{g}		&	M4			&	L4.5+L4.5	&	18.6	&	180	&	0.09		&	0.8		&	1	&	10	&	0.25	&	0.07	&				17,4,6\\
eta Tel\tablenotemark{g}			&	A0V			&	L1		&	4.2	&	190	&	$<$0.15	&	$<$7		&	0.08	&	0.02	&	2.08	&	0.20	&				8,37\\
GSC 08047-00232\tablenotemark{g}&	K3			&	M9.5		&	3.2	&	200	&	$<$0.1	&	$<$7		&	0.01	&	0.04	&	0.50	&	0.02	&		16,19\\
GG Tau\tablenotemark{g}			&	K7+M0.5+M5+	&	M7		&	1.5	&	210	&	$<$0.1	&	$<$14	&	0.01	&	0.02	&	1.10	&	0.04	&	5\\
2MASS J0551-4434				&	M8.5			&	L0		&	2.2	&	220	&	$<$1		&	$<$62	&	0.1	&	10	&	0.07	&	0.06	&		23\\
LP 213-67\tablenotemark{g}		&	M6.5			&	M8+L0	&	14	&	230	&	0.12		&	3		&	---	&	---	&	0.10	&	0.18\tablenotemark{e}	&			9,15\\
GJ 1048						&	K2			&	L1		&	11.9	&	250	&	$<$1		&	$<$26	&	0.6	&	2	&	0.84	&	0.07	&				10\\
HD 65216	\tablenotemark{g}		&	G5			&	M7+L2	&	7	&	253	&	0.17		&	6		&	3	&	6	&	0.94	&	0.09	&				31\\
AB Pic\tablenotemark{g}			&	K2			&	L1		&	5.5	&	275	&	$<$0.1	&	$<$6		&	$\sim$0.03	&	$\sim$0.03	&	0.70	&	0.01	&				21,38\\
G196-3						&	M2.5			&	L2		&	16.2	&	300	&	$<$1		&	$<$2	0	&	0.06	&	0.3	&	0.30	&	0.04	&				2\\
BD+13 1727					&	K5			&	M8		&	10.5	&	380	&	$<$1		&	$<$45	&	---	&	---	&	1.20	&	---	&			28\\
Wolf 940						&	M4			&	T8.5		&	32	& 	400	&	$<$1		&	$<$62	&	3.5	&	6	&	0.20	&	0.03	&				35\\
V1428 Aql\tablenotemark{f}		&	M3			&	M8		&	75	&	400	&	$<$1		&	$<$6		&	---	&	---	&	0.40	&	---	&					1\\
Denis-P J1347-7610			&	M0			&	L0		&	16.8	&	418	&	$<$1		&	$<$62	&	0.2	&	1.4	&	0.60	&	---	&	33\\
LP 655-23					&	M4			&	M8		&	20	&	450	&	$<$1		&	$<$29	&	1	&	8	&	0.26	&	0.09	&			30\\
LP 261-75					&	M4.5			&	L6		&	13	&	450	&	$<$1		&	$<$60	&	0.1	&	0.2	&	0.22	&	0.02	&			27\\
HD 3651						&	K0			&	T7.5		&	43	&	480	&	$<$1		&	$<$44	&	0.7	&	4.7	&	0.80	&	0.03	&				29\\
HD 203030					&	G8			&	L7.5		&	11	&	487	&	$<$1		&	$<$51	&	0.13	&	0.4	&	0.97	&	0.02	&			26\\
G216-7\tablenotemark{g}			&	M3.5+M3.5	&	M9.5		&	33.6	&	634	&	$<$0.3	&	$<$6		&	1	&	10	&	1.00	&	0.07	&		11\\
HN Peg\tablenotemark{g}			&	G0			&	T2.5		&	43	&	795	&	$<$0.4	&	$<$5.5	&	0.1	&	0.5	&	1.09	&	0.02	&			29\\
Gl 337\tablenotemark{g}			&	G8+K1		&	L8+L8/T	&	43	&	880	&	0.53		&	10.9		&	0.6	&	3.4	&	1.74	&	0.04	&			12,24\\
Gl 618.1						&	M0			&	L2.5		&	35	&	1090	&	$<$1		&	$<$38	&	0.5	&	12	&	0.67	&	0.06	&				12\\
eps Indi\tablenotemark{g}			&	K5			&	T1+T6	&	402	&	1460	&	0.62		&	2.2		&	0.8	&	2	&	0.67	&	0.04	&				13\\
G124-62\tablenotemark{g}		&	dM4.5e		&	L1+L1	&	44	&	1496	&	0.42		&	14.3		&	0.5	&	0.8	&	0.21	&	0.07	&			22\\
Gl 570\tablenotemark{g}			&	K4+M1.5+M3	&	T7 		&	258	&	1500	&	$<$0.1	&	$<$0.6	&	2	&	5	&	0.95	&	0.05	&	7\\
LEHPM 494					&	M6.0			&	M9.5		&	78	&	1800	&	$<$1		&	$<$62	&	2	&	10	&	0.10	&	0.08	&			30\\
Gl 417\tablenotemark{g}			&	G0+G0		&	L4.5+L6	&	90	&	2000	&	0.07		&	1.5		&	0.08	&	0.3	&	0.94	&	0.04	&			11,14\\
APMPM J2354-3316			&	DA+M4		&	M8.5		&	8	&	2200	&	$<$1		&	$<$62	&	1.8	&	1.8	&	0.65	&	0.10	&18\\
HD 89744						&	F7			&	L0		&	63	&	2460	&	$<$1		&	$<$40	&	1.5	&	3	&	1.40	&	0.07	&				12\\
Gl 584						&	G1+G3		&	L8		&	194	&	3600	&	$<$1		&	$<$20	&	1	&	2.5	&	1.10	&	0.06	&			31\\
2MASS J0126-5022				&	M6.5			&	M8		&	82	&	5100	&	$<$1		&	$<$62	&	0.2	&	2	&	0.10	&	0.09	&			34\\
2MASS J1258+4013			&	M7			&	M6		&	63	&	6700	&	$<$1		&	$<$104	&	1	&	5	&	0.11	&	0.09	&				36\\
HD 221356\tablenotemark{g}		&	F8			&	M8+L3	&	452	&    11900&	0.57		&	15		&	5.5	&	8	&	1.02	&	0.09	&		30\\

\enddata
\tablerefs{1=	\citet{1944AJ.....51...61V}
2	=	\citet{1998Sci...282.1309R}
3	=	\citet{1999ApJ...512L..69L}
4	=	\citet{1999ApJ...519..802K}
5	=	\citet{1999ApJ...520..811W}
6	=	\citet{1999Sci...283.1718M}
7	=	\citet{2000ApJ...531L..57B}
8	=	\citet{2000ApJ...541..390L}
9	=	\citet{2000MNRAS.311..385G}
10	=	\citet{2001AJ....121.2185G}
11	=	\citet{2001AJ....121.3235K}
12	=	\citet{2001AJ....122.1989W}
13	=	\citet{2003A&A...398L..29S}
14	=	\citet{2003AJ....126.1526B}
15	=	\citet{2003ApJ...587..407C}
16	=	\citet{2004A&A...420..647N}
17	=	\citet{2004AJ....128.1733G}
18	=	\citet{2004MNRAS.347..685S}
19	=	\citet{2005A&A...430.1027C}
20	=	\citet{2005A&A...435L..13N}
21	=	\citet{2005A&A...438L..29C}
22	=	\citet{2005A&A...440..967S}
23	=	\citet{2005A&A...440L..55B}
24	=	\citet{2005AJ....129.2849B}
25	=	\citet{2005AN....326..974S}
26	=	\citet{2006ApJ...651.1166M}
27	=	\citet{2006PASP..118..671R}
28	=	\citet{2007AJ....133..439C}
29	=	\citet{2007ApJ...654..570L}
30	=	\citet{2007ApJ...667..520C}
31	=	\citet{2007MNRAS.378.1328M}
32	=	\citet{2008ApJ...689..471R}
33	=	\citet{2008MNRAS.383..831P}
34	=	\citet{2009ApJ...692..149A}
35	=	\citet{2009arXiv0902.1812B}
36	=	\citet{2009arXiv0903.3251R}
37     =      \citet{2002ApJ...575L..75O}
38     =      \citet{2003ApJ...599..342S} }
\tablenotetext{a}{UCD-UCD separation estimated from the discovery paper or assumed to be 1$\arcsec$, a limit which comes from the 2MASS images (\citealt{2006AJ....131.1163S})}
\tablenotetext{b}{Ages come from the cited discovery paper}
\tablenotetext{c}{Mass of the primary estimated from the mass-luminosity relation in Allen's Astrophysical Quantities (\citealt{2000asqu.book.....C}) or \citet{2005nlds.book.....R} (for M stars) unless otherwise noted}
\tablenotetext{d}{Mass of the secondary estimated from the discovery paper}
\tablenotetext{e}{The mass of the primary in this system in less than the combined mass of the two components which make up the secondary}
\tablenotetext{f}{aka VB 10}
\tablenotetext{g}{These systems are used in the multiplicity analysis along with 2MASS J0025+4759 and 2MASS J1200+2048 as the UCD secondary has been targeted with HST or AO to resolve a closely separated ($<$20AU) pair}
\end{deluxetable} 

\clearpage
\begin{deluxetable}{rrrrrrrrrrrrrrr}
\label{tab:tab1}
\tabletypesize{\scriptsize}
\tablecaption{ Astrometric Information on the Companion Candidates\label{Astro_Comp}}
\tablewidth{0pt}
\tablehead{
\colhead{Name} &
\colhead{Ref} &
\colhead{$\mu_{\alpha}$ } &
\colhead{$ \mu_{\delta}$}  &
\colhead{SpT} &
\colhead{SpT}&
\colhead{Distance}&
\colhead{$\rho$} &
\colhead{$\rho$} \\
& &
\colhead{\arcsec yr$^{-1}$} &
\colhead{\arcsec yr$^{-1}$} &
\colhead{Opt} &
\colhead{IR}&
\colhead{pc} &
\colhead{arcsec}&
\colhead{AU}\\
\colhead{(1)} &
\colhead{(2)} &
\colhead{(3)} &
\colhead{(4)} &
\colhead{(5)} &
\colhead{(6)} &
\colhead{(7)} &
\colhead{(8)} &
\colhead{(9)} \\
}
\startdata
2MASS J00034227-2822410	&	2	&	0.257	$\pm$	0.016	&	-0.145	$\pm$	0.018	&	M8		&	M8	&	26	$\pm$	3		\\
G 266-33	&		&	0.280	$\pm$	0.001	&	-0.1431	$\pm$	0.0007	& G8 	&& $	39.5	^{+	1.8	}_{-	1.6	}$	&	66	&	2610	\\
\hline
2MASS J00250365+4759191AB	&	2	&	0.312	$\pm$	0.039	&	-0.009	$\pm$	0.044	&	L4+L4\tablenotemark{a}	&	---	&	31	$\pm$	6\tablenotemark{a}	\\	
G 171-58	&		&	0.2743	$\pm$	0.0007	&	0.0112	$\pm$	0.0009	&  F8&&	$	42.2	^{+	2.0	}_{-	1.8	}$	&	218	&	9202	\\
\hline
SDSS J004154.54+134135.5	&	3,9	&	-0.174	$\pm$	0.024	&	-0.138	$\pm$	0.036	&	L0		&	---	&	31	$\pm$	6		\\	
NLTT 2274		&		&	-0.201	$\pm$	0.013	&	-0.178	$\pm$	0.013	& M4 &M4&		21	$\pm$	8				&	23	&	483	\\
\hline
SDSS J020735.60+135556.3	&	3,7	&	0.260	$\pm$	0.017	&	-0.161	$\pm$	0.018	&	L2		&	L2	&	35	$\pm$	5		\\	
G 73-26	&		&	0.262	$\pm$	0.013	&	-0.186	$\pm$	0.013	& M2 &&		26	$\pm$	10				&	73	&	2774	\\
\hline
2MASS J12003292+2048513	&	5	&	-0.159	$\pm$	0.019	&	0.232	$\pm$	0.019	&	M7		&	---	&	26	$\pm$	3		\\	
G 121-42	&		&	-0.157	$\pm$	0.013	&	0.241	$\pm$	0.013	& M4 &&		 $	30^{+	14	}_{-	7	}$	&	204	&	5916	\\
\hline
2MASS J13204159+0957506	&	6,7	&	-0.236	$\pm$	0.021	&	-0.129	$\pm$	0.021	&	M8	&	---	&	36	$\pm$	3		\\	
G 63-23	&		&	-0.250	$\pm$	0.002	&	-0.144	$\pm$	0.002	& K5&&	$	38.1	^{+	2.6	}_{-	2.3	}$	&	169	&	6445	\\
\hline
2MASS J13204427+0409045	&	6,8	&	-0.483	$\pm$	0.019	&	0.211	$\pm$	0.017	&	L3		&	---	&	33	$\pm$	3		\\	
G 62-33	&		&	-0.507	$\pm$	0.001	&	0.202	$\pm$	0.0009	& K2&&	$	30.5	^{+	1.0	}_{-	1.0	}$	&	66	&	2010	\\
\hline
SDSS J141659.78+500626.4	&	1	&	-0.297	$\pm$	0.013	&	0.188	$\pm$	0.021		&	---		&	L4	&	44	$\pm$	31		\\	
G 200-28	&		&	-0.3003	$\pm$	0.0007	&	0.1861	$\pm$	0.0007	& G5&&	$	45.1	^{+	1.6	}_{-	1.5	}$	&	570	&	25734	\\
\hline
SDSS J175805.46+463311.9	&	4	&	0.026	$\pm$	0.015	&	0.594	$\pm$	0.016	&	---		&	T6.5	&	12	$\pm$	2		\\	
G 204-39	&		&	-0.017	$\pm$	0.002	&	0.575	$\pm$	0.002	& M3&&	$	13.6	^{+	0.3	}_{-	0.3	}$	&	198	&	2685	\\
\hline

\enddata
\tablerefs{
1	=	\citet{2006AJ....131.2722C}
2	=	\citet{2007AJ....133..439C}
3	=	\citet{2002AJ....123.3409H}
4	=	\citet{2004AJ....127.3553K}
5	=	\citet{2000AJ....120.1085G}
6	=	\citet{2008AJ....136.1290R}
7       =     \citealt{2009MNRAS.394..857D}
8       =   \citet{2006MNRAS.368.1281P}
9   =      \citet{2008MNRAS.384.1399J}}

\tablenotetext{a}{This L4+L4 distance is reported in \citet{2006AJ....132..891R}}

\end{deluxetable}
\clearpage

\begin{deluxetable}{lcccccccrrr}
\label{tab:tab1}
\tabletypesize{\scriptsize}
\tablecaption{ Reliability of the Common Proper Motion Pairs\label{Reliability}}
\tablewidth{0pt}
\tablehead{
\colhead{Name} &
\colhead{Num Match\tablenotemark{a}}  &
\colhead{\% Chance Alignment } &
\colhead{Num Match\tablenotemark{a}} &
\colhead{\% Chance Alignment } \\
 &
\colhead{LSPM-N} &
\colhead{LSPM-N} &
\colhead{Hipparcos} &
\colhead{Hipparcos} \\

\colhead{(1)} &
\colhead{(2)} &
\colhead{(3)} &
\colhead{(4)} \\
}
\startdata

2MASS J0003-2822& 			259 &	0.01	&	63 &0.01 			\\
2MASS J0025+4759 &			283	&	0.85	&	68 &0.02			\\
SDSS J0041+1341 &			2336 &	0.44	&	294&$<$0.01		\\
SDSS J0207+1355 &			230	&	0.04	&	49&	$<$0.01		\\
2MASS J1200+2048 &			55 &		0.07	&	15 &$<$0.01		\\
2MASS J1320+0957 &			418	&	1.03	&	73&	0.02			\\
2MASS J1320+0409 &			11	&	$<$0.01&	6 &$<$0.01	\\
SDSS J1416+5006 &			40 &		0.12	&	13&	0.01			\\
SDSS J1758+4633 &			2  &		$<$0.01&	3& $<$0.01	\\
\hline
\enddata
\tablenotetext{a}{These columns tabulate the number of stars in the entire Hipparcos or LSPM-N catalog that had matching proper motion components to the UCD at the 2$\sigma$ level}

\end{deluxetable}
\clearpage

\begin{deluxetable}{lcccrrrrrrrrrrr}
\label{tab:tab1}
\tabletypesize{\scriptsize}
\tablecaption{ Details of SMARTS Observations\label{SMARTS}}
\tablewidth{0pt}
\tablehead{
\colhead{Name} &
\colhead{Instrument} &
\colhead{Exposure Time} &
\colhead{Date} &
\colhead{Airmass}  &
\colhead{Grating}  \\
 & &
\colhead{(s)} &
&
 
 &
 &
 \\

\colhead{(1)} &
\colhead{(2)} &
\colhead{(3)} &
\colhead{(4)} &
\colhead{(5)} &
\colhead{(6)}\\
}
\startdata
G 266-33		&R-C Spec	&2100		&	12 November 2008	&1.001&47/II\\	
G 266-33		&R-C Spec	&1800		&	30 November 2008	&1.313&26/Ia	\\		
G 266-33		&Echelle		&1500		&	20 April 2009	&1.370&---	\\
\hline
NLTT 2274	&R-C Spec	&2700		&	29 September 2008	&1.571&47/Ib	\\
NLTT 2274	&R-C Spec	&1800		&	24 October 2008	&1.387&26/Ia	\\
NLTT 2274	&R-C Spec	&2700		&	17 November 2008	&1.413&26/Ia	\\
\hline
G 73-26	&R-C Spec	&1800		&	17 September 2008	&1.883&47/Ib	\\
G 73-26	&R-C Spec	&1800		&	02 October 2008	&1.812&26/Ia	\\
G 73-26	&R-C Spec	&1800		&	16 November 2008	&1.438&47/Ib	\\
G 73-26	&R-C Spec	&2700		&	26 November 2008	&1.393&47/II	\\
\hline
G 121-42	&R-C Spec	&1500		&	25 December 2008	&2.026&32/I	\\
G 121-42	&R-C Spec	&1800		&	25 January 2009	&1.766&26/Ia	\\
G 121-42	&R-C Spec	&1500		&	25 December 2009	&1.420&47/Ib	\\
G 121-42	&R-C Spec	&1200		&	25 December 2009	&1.827&47/II	\\

\hline
G 63-23		&Echelle		&1800		&	14 November 2008	&1.013&---	\\
G 63-23		&R-C Spec	&1800		&	29 January 2009	&1.647&26/Ia	\\
G 63-23		&R-C Spec	&1200		&	14 February 2009	&1.669&47/Ib	\\
G 63-23		&R-C Spec	&3600		&	24 February 2009	&1.306&47/II	\\
\hline
G 62-33		&R-C Spec	&1200		&	29 January 2009	&1.637&26/Ia	\\
G 62-33		&R-C Spec	&1200		&	14 February 2009	&1.660&47/Ib	\\
G 62-33		&Echelle		&1500		&	21 February 2009	&1.262&---	\\
G 62-33		&R-C Spec	&1800		&	24 February 2009	&1.223&47/II	\\

\hline
\hline
\enddata
\tablecomments{Grating 26/Ia covers 3700-5400~\AA\ at 4.4~\AA\ spectral resolution, Gratings 47/Ib and 47/II cover 5600-6950~\AA\ at 3.1~\AA\ spectral resolution}

\end{deluxetable}

\clearpage

\begin{deluxetable}{lcccrrrrrrrrrrr}
\label{tab:tab1}
\tabletypesize{\scriptsize}
\tablecaption{ Details of KPNO Observations\label{KPNO}}
\tablewidth{0pt}
\tablehead{
\colhead{Name} &
\colhead{Instrument} &
\colhead{Exposure Time} &
\colhead{Date} \\
 & &
\colhead{(s)} &
&
 
 &
 &
 \\

\colhead{(1)} &
\colhead{(2)} &
\colhead{(3)} &
\colhead{(4)} \\
}
\startdata
G 62-33		&Echelle	&	915	&26,27 June 2009			\\	
G 63-23		&Echelle	&	500	&27 June 20009			\\	
G 200-28		&Echelle	&	900	&25 June 2009				\\	
G 171-58		&Echelle	&	900	&25 June 2009				\\	

\hline
\hline
\enddata
\end{deluxetable}

\begin{deluxetable}{lcrrrrrrrrrrrrr}
\label{tab:tab1}
\tabletypesize{\scriptsize}
\tablecaption{ Details of MagE Observations\label{MagE}}
\tablewidth{0pt}
\tablehead{
\colhead{Name} &
\colhead{Exposure Time} &
\colhead{Date} &
\colhead{Airmass} \\
 &
\colhead{(s)} &
&
&
\\

\colhead{(1)} &
\colhead{(2)} &
\colhead{(3)} &
\colhead{(4)}  \\
}
\startdata
G 266-33& 5 & 25 November 2008 & 1.227   \\ 			
2MASS J0003-2822& 1200 & 25 November 2008 & 1.266  \\ 	
NLTT 2274 &100& 07 October 2008 &1.443 \\					
2MASS J0041+1341 &1500& 07 October 2008 & 1.464  \\ 				
G 73-26 &120& 25 November 2008 & 1.463 \\ 				
2MASS J0207+1355 &2400 & 08 October 2008 & 1.376 \\ 				
G 121-42 &100 & 07 March 2009 & 1.111 \\
G 62-33&10&11 January 2009&2.038\\				
G 63-23&30&11 January 2009&2.264\\
\hline
\hline
\enddata
\end{deluxetable}

\begin{deluxetable}{lcccrrrrrrrrrrr}
\label{tab:tab1}
\tabletypesize{\scriptsize}
\tablecaption{ Details of SpeX Observations\label{SpeX}}
\tablewidth{0pt}
\tablehead{
\colhead{Name} &
\colhead{Exposure Time} &
\colhead{Date} &
\colhead{Airmass} &
\colhead{Calibration Star} \\

 &
\colhead{(s)} &
&
 &
 &
 &
 \\

\colhead{(1)} &
\colhead{(2)} &
\colhead{(3)} &
\colhead{(4)} &
\colhead{(5)} \\
}
\startdata
2MASS J0003-2822& 450 &   09 December 2008 & 1.499 & HD 220455 \\ 	
2MASS J0025+4759& 510   & 09 December 2008 & 1.186 & HD 1561\\		
NLTT 2274 & 360    & 09 December 2008 & 1.006 & HD 6457 \\ 		
G 73-26 & 360    &15 December 2008 & 1.070& BD+18 337A \\ 
2MASS J0207+1355 & 510   & 10 December 2008 & 1.037 & V* Vz ari\\ 				
\hline
\hline
\enddata
\end{deluxetable}

\begin{deluxetable}{lcccccccc}
\label{tab:tab1}
\tabletypesize{\scriptsize}
\tablecaption{ Details of ANDICAM Observations\label{PHOTO}}
\tablewidth{0pt}
\tablehead{
\colhead{Name} &
\colhead{Date} &
\colhead{Number of Images} &
\colhead{Band} \\

 &
\colhead{(s)} &
&
 &
 &
 &
 \\

\colhead{(1)} &
\colhead{(2)} &
\colhead{(3)} &
\colhead{(4)} \\
}
\startdata
G 62-33  &16 February 2009 -- 9 April 2009  &54&$B$ \\
G 62-33  & 16 February 2009 -- 26 March 2009 &30&$V$ \\
G 62-33  & 27 March 2009 -- 9 April 2009  &24&$I$ \\
\hline
G 63-23  &10 February 2009 -- 3 April 2009 &32 &$V$\\
G 63-23  &10 February 2009 -- 3 April 2009 & 32&$I$\\
\hline
G 121-42  &10 February 2009 -- 31 May 2009  & 46 &$V$\\ 
G 121-42  &10 February 2009 -- 31 May 2009  & 46 &$I$\\ 
\hline
G 73-26  &4 December 2008 -- 31 January 2009 &31 &$V$\\
G 73-26  &4 December 2008 -- 31 January 2009 &31 &$I$\\
\hline
\hline
\enddata
\end{deluxetable}

\begin{deluxetable}{rrrrrrrrrrrrrrrr}
\label{tab:tab1}
\tabletypesize{\scriptsize}
\rotate
\tablecaption{ Details of the Primaries\label{Primaries}}
\tablewidth{0pt}
\tablehead{
\colhead{Name} &
\colhead{SpT} &
\colhead{[Fe/H] } &
\colhead{log($R'_{HK}$)}  &
\colhead{U}  &
\colhead{V}  &
\colhead{W}  &
\colhead{W$_{\lambda}$(Li)}  &
\colhead{W$_{\lambda}$(H$\alpha$)}\tablenotemark{a}   &
\colhead{L$_{H\alpha}$/L$_{bol}$}\tablenotemark{b}   &
\colhead{Mass}  &
\colhead{$P_{rot}$}  &
\colhead{Member?}  &
\colhead{Age}  &
\colhead{Ref}  \\

&
&
&
&
\colhead{(km s$^{-1}$)} &
\colhead{(km s$^{-1}$)} & 
\colhead{(km s$^{-1}$)} &
\colhead{(\AA)}&
\colhead{(L$_{\sun}$)}&
\colhead{(\AA)}&
\colhead{($M_{\sun}$)}&
\colhead{(days)}&

&
\colhead{(Gyr)}&
&\\
\colhead{(1)}&
\colhead{(2)}&
\colhead{(3)}&
\colhead{(4)}&
\colhead{(5)}&
\colhead{(6)}&
\colhead{(7)}&
\colhead{(8)}&
\colhead{(9)}&
\colhead{(10)}&
\colhead{(11)}&
\colhead{(12)}&
\colhead{(13)}&
\colhead{(14)}&
\colhead{(15)}\\

}
\startdata
G 266-33 		&G8		&0.07, 0.097	&-4.55	&-32		&-47	&-20	&$<$0.004&-1.20$\pm$0.09	&---		&0.94&---&---			&0.9-1.4	&1, 2,3,4,11,12 	\\
G 171-58		&F8		&0.22		&-4.81		&-48		&-26	&-4	&---		&---				&---		&1.15&---&---			&1.8-3.5	&1,4,5,11			\\
NLTT 2274	&M4		&---			&---		&---		&---	&---	&$<$0.03	&$>$0.10			&$<$-5.20&0.20&---&---			&4.5-10	& 9,12			\\
G 73-26		&M2		&---			&---		&-44		&-88	&68	&$<$0.04	&$>$0.40			&$<$-4.41&0.44&39.6$\pm$0.6&---	&3-4		& 9,12			\\
G 121-42		&M4		&---			&---		&---		&---	&---	&$<$0.40	&$>$0.20			&$<$-4.71&0.20&47.0$\pm$0.9&---	&4-5		& 9,10,12			\\
G 62-33		&K2		&-0.18,0.15	&-4.77	&-73		&-21	&21	&$<$0.004&-1.0$\pm$0.10	&---		&0.85&---&---			&3.3-5.1		&1,4,6,11,12		\\
G 63-23		&K5		&---			&-4.49	&-20		&-50	&6	&$<$0.006&-0.80$\pm$0.05	&---		&0.67&---&---			&0.5-3	& 4,11,12			\\
G 200-28		&G5		&-0.16		&$<$-5.00&-72		&-40	&-35	&---		&---				&---		&1.01&---&---			&7-12	& 1,4,11			\\
G 204-39		&M3		&---			&---		&-35		&8	&8	&---		&-0.215			&---		&0.36&---&Hyades SC	&0.5-3	&4,7,8,11			\\
\hline
\enddata
\tablenotetext{a}{H$\alpha$ EW is given reported with (-) indicating absorption and (+) indicating emission}
\tablenotetext{b}{ ~The L$_{H\alpha}$/L$_{bol}$ quantity is calibrated as an age/activity indicator for M dwarfs and not for higher temperature stars}
\tablecomments{ References: 1=\citet{2008yCat.5128....0H}  2=\citet{1998A&A...339..791R}   3=\citet{1996AJ....111..439H}     4= \citet{2006A&AT...25..145G}  5=\citet{2005yCat.5122....0P}    6=\citet{2002yCat..33940927I}  7=\citet{2002AJ....123.3356G}   8=\citet{1993AJ....106.1885E,1990PASP..102..166E}  9=\citet{2002AJ....124.1190L} 10=\citet{1995gcts.book.....V} 11=\citet{1997A&A...323L..49P} 12=This paper}

\end{deluxetable}

\begin{deluxetable}{ccccccccccc}
\label{tab:tab1}
\tabletypesize{\scriptsize}
\rotate
\tablecaption{ Details of the UCD Secondaries\label{Secondaries}}
\tablewidth{0pt}
\tablehead{
\colhead{Name} &
\colhead{SpT} &
\colhead{log(L$_{bol}$)} &
\colhead{W$_{\lambda}$(H$\alpha$)} &
\colhead{log(L$_{H\alpha}/L_{bol}$)} &
\colhead{W$_{\lambda}$(Li)} &
\colhead{$J-K_{s}$} &
\colhead{Age} &
\colhead{Mass} &
\colhead{References} \\ & &
\colhead{(L$_{\sun}$)} &
\colhead{(\AA)} & &
\colhead{(\AA)} & &
\colhead{(Gyr)}&
\colhead{($M_{\sun}$)}  &\\
\colhead{(1)} &
\colhead{(2)} &
\colhead{(3)} &
\colhead{(4)} &
\colhead{(5)} &
\colhead{(6)} &
\colhead{(7)} &
\colhead{(8)} \\
}
\startdata
2MASS J0003-2822& 	M8		&[-2.85,-2.93]		&9.0$\pm$0.08&	 -4.27  &$<$0.18	&1.096$\pm$0.035	&0.1-1	 	&0.100-0.103& 4 \\ 
2MASS J0025+4759 &	L4/L4	&[-3.57,-3.69]	&$<$0.10		&     ---          &10.0$\pm$1.0	&1.938$\pm$0.069	&0.1-0.5			&0.045-0.065\tablenotemark{a}&	1,2,4	\\
2MASS J0025+4759&	L4/L4	&[-3.57,-3.69]	&$<$0.10		&     ---          &10.0$\pm$1.0	&1.938$\pm$0.069	&0.1-0.5			&0.080-0.083\tablenotemark{b} &	1,2,4	\\
SDSS J0041+1341 &	L0		&[-3.53,-3.85]	&2.2$\pm$0.10&---     	   &$<$0.40		&1.218$\pm$0.042	&2-8			&0.081-0.083&  4         \\	
SDSS J0207+1355 &	L3		&[-3.78,-3.95]	&$<$0.20		&            ---   &$<$0.30		&1.550$\pm$0.085	&2-8			&0.079-0.081&	 4	\\	
2MASS J1200+2048 &	M7		&[-2.86,-3.42]	&2.9			& $<$-5.44  &$<$0.7			&1.001$\pm$0.030	&5-7			&0.085-0.103&	 3,4	\\	
2MASS J1320+0409 &	L3		&[-3.85,-3.94]	&$<$0.80		&            ---   &$<$1.0			&1.625$\pm$0.065	&2-8			&0.079-0.081&	4	\\	
2MASS J1320+0957 &	M8		&[-3.12,-3.23]	&$<$0.40		& $<$-5.74  &$<$0.30		&1.117$\pm$0.039	&1-8			&0.083-0.093&	4 \\	
SDSS J1416+5006 &	L4		&[-4.20,-4.31]	&---			&		---&             ---      	&1.560$\pm$0.085	&---				&0.077-0.078&	4	\\
SDSS J1758+4633 &	T6.5		&[-5.12,-5.24]	&---			&		---&              ---     	&0.180$\pm$0.085	&0.5-1.5			&0.020-0.035&4	\\
\hline
\enddata	
\tablenotetext{a}{Calculated using the age from diagnostics of the secondary}
\tablenotetext{b}{Calculated using the age from diagnostics of the primary}
\tablecomments{ References:  1= \citet{2006AJ....132..891R}  2=\citet{2007AJ....133..439C}  3=\citealt{2002AJ....123.2806R} 4=This paper}
\end{deluxetable}
\begin{deluxetable}{cccccc}
\label{tab:tab1}
\tabletypesize{\tiny}
\tablecaption{ Estimated Ages of the Systems\label{Ages}}
\tablewidth{0pt}
\tablehead{
\colhead{Name} &
\colhead{Name} &
\colhead{Age} &
\colhead{Age}  &
\colhead{Age}  \\
\colhead{(primary)} &
\colhead{(secondary)} &
\colhead{(primary)} &
\colhead{(secondary)}&
\colhead{(system)}\\
& &
\colhead{(Gyr)}&
\colhead{(Gyr)}&
\colhead{(Gyr)}\\
\colhead{(1)} &
\colhead{(2)} &
\colhead{(3)} &
\colhead{(4)}&
\colhead{(5)}\\
}
\startdata
G 266-33 			&2MASS J0003-2822		& 	0.9-1.4		&	0.1-1.0	&0.9-1.4&	\\
G 171-58			&2MASS J0025+4759 		&	1.8-3.5		&	0.1-0.5	&----&	\\
NLTT 2274		&2MASS J0041+1341 		&	4.5-10		&	2-8		&4.5-8&\\
G 73-26			&2MASS J0207+1355 		&	3-4			&	2-8		&3-4&\\
G 121-42			&2MASS J1200+2048 		&	4-5			&	5-7		&4-5&\\
G 62-33			&2MASS J1320+0409 		& 	3.3-5.1		&	2-8		&3.3-5.1&\\
G 63-23			&2MASS J1320+0957 		&	1.0-3			&	1-8		&1.0-3&\\
G 200-28			&2MASS J1416+5006 		&	7-12			&	---		&7-12&\\
G 204-39			&2MASS J1758+4633 		&	0.5-3			&	0.5-1.5	&0.5-1.5&	\\
\hline
\enddata
\end{deluxetable}

\clearpage
\bibliographystyle{apj}
\bibliography{ms_RR}

\end{document}